\documentclass[sigconf, screen]{acmart}
\AtBeginDocument{%
  }

\setcopyright{acmlicensed}
\copyrightyear{2025}
\acmYear{2025}
\setcopyright{cc}
\setcctype{by}
\acmConference[CCS '25]{Proceedings of the 2025 ACM SIGSAC Conference on Computer and Communications Security}{October 13--17, 2025}{Taipei, Taiwan}
\acmBooktitle{Proceedings of the 2025 ACM SIGSAC Conf. on Computer and Communications Security (CCS '25), Oct. 13--17, 2025, Taipei, Taiwan}
\acmDOI{10.1145/3719027.3765160}
\acmISBN{979-8-4007-1525-9/2025/10}

\usepackage{xspace}
\usepackage{tikz}
\usepackage [
lambda,
advantage,
operators,
sets,
landau,
probability,
notions,
logic,
ff,
mm,
primitives,
events,
complexity,
asymptotics,
keys
]{cryptocode}
\usepackage{verbatim}
\usepackage{pgfplots}
\usepgfplotslibrary{statistics}
\usetikzlibrary{math} 
\pgfplotsset{width=\columnwidth,compat=1.9} 
\usepackage{amsmath}
\usepackage{algorithm}
\usepackage{algpseudocode}
\usepackage{multirow}
\usepackage{numprint}
\usepackage{enumitem}
\usepackage{url}
\usepackage{placeins}
\usepackage{needspace}

\newcommand\sysNoSpace{$\mathcal{Z}$ORRO}
\newcommand\sysGen{\texttt{\sysNoSpace}'s\xspace}
\newcommand\sys{\texttt{\sysNoSpace}\xspace}
\newcommand{\ourname}{\sys}
\newcommand{\ournameGen}{\sysGen}

\newcommand{\paperTitle}{\sysNoSpace: Zero-Knowledge Robustness and Privacy \\ for Split Learning (Full Version)}\newcommand{\paperTitleNoLineBreak}{\sysNoSpace: Zero-Knowledge Robustness and Privacy for Split Learning (Full Version)}

\newcommand{\Prv}{$\mathcal{P}$\xspace}
\newcommand{\Vrf}{$\mathcal{V}$\xspace}
\newcommand{\Cir}{$\mathcal{C}$\xspace}

\newcommand{\queueLength}{\ensuremath{k}\xspace}
\newcommand{\betaParam}{\ensuremath{\beta}\xspace}
\newcommand{\bestModel}{\textit{BM}\xspace}

\newcommand{\cifar}{\mbox{CIFAR-10}\xspace}
\newcommand{\cifarhundred}{\mbox{CIFAR-100}\xspace}
\newcommand{\tinyimagenet}{\mbox{Tiny ImageNet}\xspace}
\newcommand{\forest}{\mbox{Forest Cover Type}\xspace}
\newcommand{\mnist}{MNIST\xspace}
\newcommand{\fmnist}{FMNIST\xspace}

\newcommand{\circled}[1]{\tikz[baseline=(char.base)]{
            \node[shape=circle,draw,inner sep=1pt] (char) {#1};}}

\newcommand{\etal}{et al.\xspace}

\newcommand{\numberOfMaliciousClients}{\ensuremath{N_{\adversary}}\xspace}

\newcommand{\scaleTable}[1]{\scalebox{0.75}{#1}}

\newcommand{\goldenDefense}{Gold Standard\xspace}

\newcommand{\sect}{Sect.~}

\newcommand{\app}{App.~}

\newcommand{\numberOfClients}{N\xspace}

\newcommand{\adversaryNoSpace}{\ensuremath{\mathcal{A}}}
\newcommand{\adversary}{\adversaryNoSpace\xspace}

\createprocedureblock{interactGame}{center,boxed}{}{}{}
\newcommand{\cellbreaks}[1]{\begin{tabular}[c]{@{}c@{}}#1\end{tabular}}

\NewDocumentEnvironment{gameInteract}{m+b}
 {%
  \interactGame[linenumbering]{#1}{#2}
  \par\addvspace{\baselineskip}
 }{}

\algnewcommand{\Input}[1]{\State \textbf{Input:} #1}
\algnewcommand{\Output}[1]{\State \textbf{Output:} #1}

\newcommand{\changed}[1]{#1}
\newcommand{\changedRevision}[2]{#2}
\newcommand{\changedEditorial}[1]{#1}

\begin{document}

\title[\paperTitleNoLineBreak]{\paperTitle}

\author{Nojan Sheybani}
\email{nsheyban@ucsd.edu}
\authornote{These authors contributed equally to this work.}
\affiliation{%
  \institution{University of California San Diego}
  \city{La Jolla}
  \country{USA 
  }
}
\author{Alessandro Pegoraro}
\email{alessandro.pegoraro@trust.tu-darmstadt.de}
\authornotemark[1]
\affiliation{%
  \institution{Technical University of Darmstadt}
  \city{Darmstadt}
  \country{Germany}
}
\author{Jonathan Knauer}\authornotemark[1]
\email{jonathan.knauer@stud.tu-darmstadt.de}
\affiliation{
  \institution{Technical University of Darmstadt}
  \city{Darmstadt}
  \country{Germany}
}
\author{Phillip Rieger}
\email{phillip.rieger@trust.tu-darmstadt.de}
\affiliation{%
  \institution{Technical University of Darmstadt}
  \city{Darmstadt}
  \country{Germany}
}
\author{Elissa Mollakuqe}
\email{elissa.mollakuqe@tu-darmstadt.de}
\affiliation{%
  \institution{Technical University of Darmstadt}
  \city{Darmstadt}
  \country{Germany}
}
\author{Farinaz Koushanfar}
\email{fkoushanfar@ucsd.edu}
\affiliation{%
  \institution{University of California San Diego}
  \city{La Jolla}
  \country{USA}
}

\author{Ahmad-Reza Sadeghi}
\email{ahmad.sadeghi@trust.tu-darmstadt.de}
\affiliation{%
  \institution{Technical University of Darmstadt}
  \city{Darmstadt}
  \country{Germany}
}


\renewcommand{\shortauthors}{Nojan Sheybani, Alessandro Pegoraro, Jonathan Knauer, Phillip Rieger, Elissa Mollakuqe, Farinaz Koushanfar, and Ahmad-Reza Sadeghi}
\renewcommand{\shortauthors}{Nojan Sheybani et al.}

\begin{abstract}

    Split Learning (SL) is a distributed learning approach that enables resource-constrained clients to collaboratively train deep neural networks (DNNs) by offloading most layers to a central server while keeping in- and output layers on the client-side. This setup enables SL to leverage server computation capacities without sharing data, making it highly effective in resource-constrained environments dealing with sensitive data. However, the distributed nature enables malicious clients to manipulate the training process. By sending poisoned intermediate gradients, they can inject backdoors into the shared DNN. Existing defenses are limited by often focusing on server-side protection and introducing additional overhead for the server. A significant challenge for client-side defenses is enforcing malicious clients to correctly execute the defense algorithm.

    We present \sys, a private, verifiable, and robust SL defense scheme. Through our novel design and application of interactive zero-knowledge proofs (ZKPs), clients prove their correct execution of a client-located defense algorithm, resulting in proofs of computational integrity attesting to the benign nature of locally trained DNN portions. Leveraging the frequency representation of model partitions enables \sys to conduct an in-depth inspection of the locally trained models in an untrusted environment, ensuring that each client forwards a benign checkpoint to its succeeding client. In our extensive evaluation, covering different model architectures as well as various attack strategies and data scenarios, we show \ournameGen effectiveness, as it reduces the attack success rate to less than 6\% while causing even for models storing \numprint{1000000} parameters on the client-side an overhead of less than 10 seconds.

\end{abstract}





\maketitle

\section{Introduction}
Due to its ability to succeed in increasingly complex tasks, Deep Learning has been heavily applied in ubiquitous computing systems, including domains with sensitive data. A key challenge in such applications is the availability of sufficient training data, especially in times of rising privacy concerns and regulations such as the GDPR~\cite{GDPR2018}, HIPAA~\cite{HIPAA1996}, and CCPA~\cite{CCPA2018}. To address this, in the past, schemes such as Federated Learning have been developed where the training process is outsourced to clients holding the data~\cite{mcmahan2017}. However, as model sizes continue to grow, reaching hundreds of billions of parameters~\cite{meta2025llama4}, many data holders lack the computational resources required for training while also being unable or unwilling to share their data.

Split Learning (SL) offers a promising solution. It is a distributed learning paradigm that allows clients to utilize external computational resources without exposing their data. In SL, the DNN is divided into partitions distributed across different entities~\cite{gupta2018distributed}. In the commonly used U-shaped configuration, the initial and final layers reside on the client side, while the computationally intensive middle layers are processed on a powerful server. During training and inference, hidden representations (forward pass) and gradients (backward pass) are exchanged only at partition boundaries~\cite{yang2022robust, vepakomma2018split}. By rotating the server-side model component among clients, SL facilitates collaborative training while maintaining data privacy~\cite{tajalli2023feasibility}. This architecture not only enables efficient training and inference for large DNNs, but also significantly mitigates the risk of inference attacks, as clients do not share the model components responsible for sensitive feature extraction.

\noindent\textbf{Backdoor Attacks and Defenses.} While splitting the model between clients and servers improves both privacy and computational efficiency, it also introduces a critical limitation since no single party has access to the full model, making it difficult to detect potential manipulations. Recent work has explored backdoor attack vectors, originating either from the server side~\cite{yu2024chronic, tajalli2023feasibility, pu2024dullahan} or the client side~\cite{yu2023backdoor, he2023backdoor, bai2023villain}. Client-side attacks are particularly concerning, as adversarial clients, especially in mobile or distributed settings, can operate anonymously and face minimal reputational risk if detected. In contrast, the server is typically identifiable and has an intrinsic motivation to \mbox{maintain its reputation by producing reliable models.}

Only a few approaches have been made to mitigate backdoor attacks in SL. Pu \etal consider a malicious server that aims to inject a backdoor~\cite{pu2024dullahan}. Only limited attention is given to client attacks~\cite{bai2023villain}. Rieger et al. introduced SafeSplit, a server-side defense mechanism designed to mitigate backdoor threats in SL~\cite{rieger25safesplit}. However, such server-centric defenses often impose significant computational overhead for the server and face scalability challenges, particularly when operations as pairwise distance computations are involved.

\noindent\textbf{Goals and Contributions.} We introduce \ourname, the first client-side defense mechanism designed specifically to protect SL from backdoor attacks performed by malicious clients. To understand our approach intuitively, consider how drivers place a warning triangle after an accident to alert others, allowing them to bypass danger safely. Similarly, in \ourname, each client proactively examines its own trained model segments and prevents sharing if it detects signs of poisoning. Built on zero-knowledge proofs (ZKPs) to verify the correct defense execution, \ourname forces attackers into a dilemma. They must either follow our defense protocol and eliminate their malicious updates or deviate and \mbox{consequently fail the verification step.}

Unlike previous solutions, which primarily rely on the central server for security checks, we empower clients themselves to carry out detailed inspections of their locally trained models. By leveraging frequency-domain analysis, a technique effective in revealing subtle manipulations in DNNs, we can detect poisoned model updates regardless of how many clients are compromised. To preserve model integrity and limit the propagation of poisoned contributions, each client forwards only a subset of the most recent model checkpoints to the succeeding client. The local inspection process ensures that poisoned models are filtered before being shared, mitigating the impact of backdoor attacks. At the same time, this approach minimizes the number of clients that gain visibility into any single model update, contributing to stronger privacy preservation within the collaborative training process.

To guarantee trustworthy execution of these checks without exposing sensitive data, we employ interactive zero-knowledge proofs (ZKPs). They allow clients (the "provers") to convincingly demonstrate to the server (the "verifier") that they have executed the defense protocol correctly without revealing private details such as model parameters. In our SL setup, this means clients prove that their locally trained models have been checked for poisoning and processed according to the defense algorithm, while the server can verify these proofs without learning anything about the clients' data or models. Specifically, we use ZKPs based on vector oblivious linear evaluation (VOLE), enabling robust verification and strong security guarantees even in environments where malicious participants are present. A modular approach is taken to build \ourname to ensure that the scalable and efficient implementation can be generalized to new client-based SL defense schemes. Every part of the implementation can be swapped to achieve different goals in an efficient manner, such as replacing the $\ell_1$-norm module with an $\ell_2$-norm module.

Our main contributions are as follows:
\begin{itemize}
    \item We propose \ourname, the first client-side backdoor defense against client-side backdoor attacks in SL. Built on zero-knowledge proofs (ZKPs), we create a dilemma for attackers, forcing them to reveal themselves and warn other clients or risk being dropped from the training pipeline altogether. Unlike prior work, \ourname delegates not only the training of the input and output layers but also the defense mechanism itself to the clients. This enables a fine-grained inspection of local training contributions while preserving the privacy of sensitive model parameters (\sect\ref{sec:sys-highlevel}).

    \item We propose a novel backdoor detection mechanism that analyzes the frequency-domain representation of model updates. This method enables the precise identification of backdoor artifacts without making assumptions about the proportion of malicious clients (\sect\ref{sec:sys-frequency}).

    \item We design an interactive, modular, VOLE-based zero-knowledge proof (ZKP) protocol that allows honest clients to verify the correct execution of the defense scheme of untrusted or potentially malicious participants, ensuring protocol compliance without revealing sensitive model parameters. Its modular design enables general verification of client-side defenses, making the protocol adaptable also to future SL defense schemes (\sect\ref{sec:sys-zkp}).
    
    \item We conduct an extensive empirical evaluation of \ourname across multiple datasets, model architectures, and data distributions. Our method is tested against both naive and defense-aware adversaries, demonstrating robust performance in realistic threat settings (\sect\ref{sec:eval}).

\end{itemize}

\section{Preliminaries}

\subsection{Split Learning}

\begin{figure}
    \centering
    \includegraphics[width=0.75\columnwidth, trim=1.5cm 3.4cm 16.6cm 5.4cm, clip]{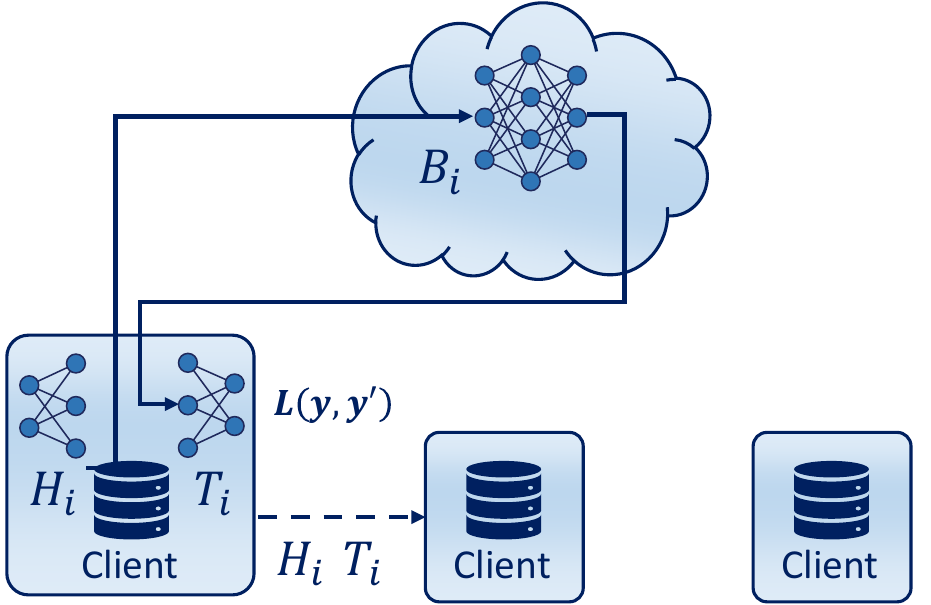}
    
    \caption{Overview of a Split Learning (SL) System.}
    \label{fig:sl}
\end{figure}

Split Learning is a collaborative machine learning framework particularly suited for scenarios where multiple clients, such as mobile devices or edge devices, hold sensitive data they do not want to share directly. In Split Learning, we assume a setting involving $N$ clients $C = {C_1, C_2, \dots, C_N}$, each of whom possesses a private dataset $Data_i$. These clients jointly train a Deep Neural Network (DNN) in coordination with a central server. A key characteristic of SL is the model-splitting approach, where the DNN is partitioned into two or three sections. The first proposed split was Vanilla Split Learning~\cite{vepakomma2018split}, with an initial model executed locally on each client and a Backbone model run on the central server. In this work, we will instead consider the U-shaped Split Learning framework~\cite{lyu2023optimal}, as depicted in Figure~\ref{fig:sl}, where each client $C_i$ computes the forward pass through a small number of initial layers (the head, denoted $H_i$), and transmits the resulting activations (smashed data) to the server. The server then continues the forward pass through the majority of the remaining layers (the backbone, denoted $B_i$) and forwards its intermediate representation back to the client which feds it to the final output layers (the tail, denoted $T_i$) to calculate the output. The client then calculates the Loss $L(y,y')$ based on the sample labels and their corresponding outputs before performing analogously the backward propagation across the model partitions. This design reduces the computational burden on edge devices and helps preserve data privacy, as the raw input never leaves the client device in both the Vanilla and U-shaped. Furthermore, the U-shaped design has the advantage of keeping the sample labels on the client device, while in the Vanilla split, they are shared with the server for backpropagation.
After the training iteration, client $C_i$ shares its Head $H_i$ and Tail $T_i$ with the next client $C_{i+1}$, as shown in Figure~\ref{fig:sl}, which uses them as the starting model for its training iteration~\cite{tajalli2023feasibility}.

\subsection{Zero-Knowledge Proofs}
\label{sec:prelims-zkp}
Zero-knowledge proofs (ZKPs) are cryptographic protocols that allow a prover \Prv to convince a verifier \Vrf of a statement's truth without revealing any information beyond its validity. These protocols are also often used for the verifiable computation of a task. In a ZKP system, \Prv demonstrates knowledge of a secret value $w$ (witness) that satisfies a computation \Cir, while \Vrf can confirm the computation's correctness without learning the secret. Formally, a ZKP system involves a prover \Prv and verifier \Vrf executing probabilistic polynomial-time algorithms that adhere to three key properties: completeness, soundness, and zero-knowledge. These properties essentially ensure that a malicious prover cannot generate a valid proof of a false statement and, no matter what, no information about the secret value is revealed.

ZKPs can be categorized as interactive or non-interactive. Interactive ZKPs require multiple communication rounds, yielding proofs verifiable only by the specific verifier involved. Non-interactive ZKPs (NIZKs) allow a prover to generate a single, publicly-verifiable proof~\cite{ben2014succinct, ben2018scalable}.
Recent NIZK advancements like zk-SNARKs~\cite{ben2014succinct} and zk-STARKs~\cite{ben2018scalable} offer compact, efficient proofs but rely on computationally intensive setups through a trusted third party or a computationally powerful verifier, hindering practicality despite widespread adoption. NIZKs also demand powerful provers for succinctness.
While lacking public verifiability, interactive ZKPs achieve significantly better efficiency and scalability, suiting computationally complex tasks. In this work, the ability to interact with two $\mathcal{V}$s in parallel makes interactive ZK an elegant solution that balances performance and communication.
\noindent\\

\textbf{Vector Oblivious Linear Evaluation (VOLE)-based ZK} protocols offer post-quantum security and efficiency through \Prv-\Vrf interaction~\cite{boyle2018compressing}. These interactive protocols achieve high efficiency and scalability using IT-MAC commitments, efficiently implemented with VOLE~\cite{baum2023sok}. Our work uses Wolverine~\cite{weng2021wolverine}, a state-of-the-art VOLE-based ZK protocol minimizing prover/verifier complexity interactively. In Wolverine, IT-MACs commit to authenticated wire values within arithmetic or boolean circuits (\Cir). The prover \Prv proves knowledge of a private vector \textbf{w}, representing the inputs, outputs, and intermediate values of \Cir. This is used to prove $\mathcal{C}(\textbf{w})=1$, while proving the consistency of protocol values \textbf{x}~\cite{weng2021wolverine}. Efficient boolean and arithmetic conversions~\cite{weng2021mystique} enable building efficient mixed-computation solutions. Due to the inherent interactivity, these protocols yield \textit{designated-verifier} proofs, limiting verification by arbitrary parties. This limitation is acceptable for SL's peer-to-peer training. Using VOLE-based ZK integrates privacy, integrity, and robustness guarantees into \sys, enabling unparalleled scalability and efficiency.
\section{Problem Setting}
\label{sec:problem}

\subsection{System Setting}
\label{sec:problem-requirements}
In the following, we consider a system consisting of \numberOfClients clients, each holding a private dataset. The clients collaboratively train a DNN without revealing their data. Due to limited local computational capabilities, the clients employ SL, coordinated by a central server having strong computational resources. Aligned with prior work~\cite{tajalli2023feasibility,rieger25safesplit}, we focus on a U-shaped SL configuration, where the head and tail of the DNN are located on the client side, while the backbone is hosted on the server.

To avoid any privacy leakage through white-box model inference attacks~\cite{he2019model,nasr2019comprehensive}, no party must have access to the entire model but only parts of it. Thus, the clients will not share trained parameters of the head or tail with the server.
\subsection{Threat Model}
\noindent\textbf{Objective:} We consider an attacker \adversary that seeks to inject a backdoor into the collaboratively trained model. The backdoor causes the model to mispredict all samples $x\in \mathcal{X_I}$ showing an adversary chosen backdoor trigger $\mathcal{I}$ as a backdoor target class $\mathcal{T}_{\adversary}$.

\noindent\textbf{Capabilities:} To perform the attack, we assume \adversary can fully control \numberOfMaliciousClients clients. Thus, \adversary can arbitrarily modify the local datasets and training procedure of these clients and also manually tamper all data the client processes, such as the head and tail. To ensure a strong adversary setting, we assume that all malicious clients know each other and are coordinated by \adversary.

\noindent\textbf{Assumptions:} 
Our assumptions are standard and aligned with existing work~\cite{rieger25safesplit}. \adversary does not control any of the benign clients, nor does it have knowledge about their local datasets. We assume that there is no collusion between malicious clients and the server, as the server's reputation would be damaged in case of an attack becoming public, while the clients can be anonymous mobile devices. Further, \adversary must avoid detection as otherwise the training process can be repeated. As such, the attack must not notably affect the model's utility. Further, an attacker is spotted and can be excluded from the training process if a deviation from the required protocol is detected, such as a failing ZKP, and this can be traced back to a specific client.  

\begin{figure*}
    \centering
    \includegraphics[width=0.9\textwidth, trim=0.5cm 1.4cm 6cm 0cm, clip]{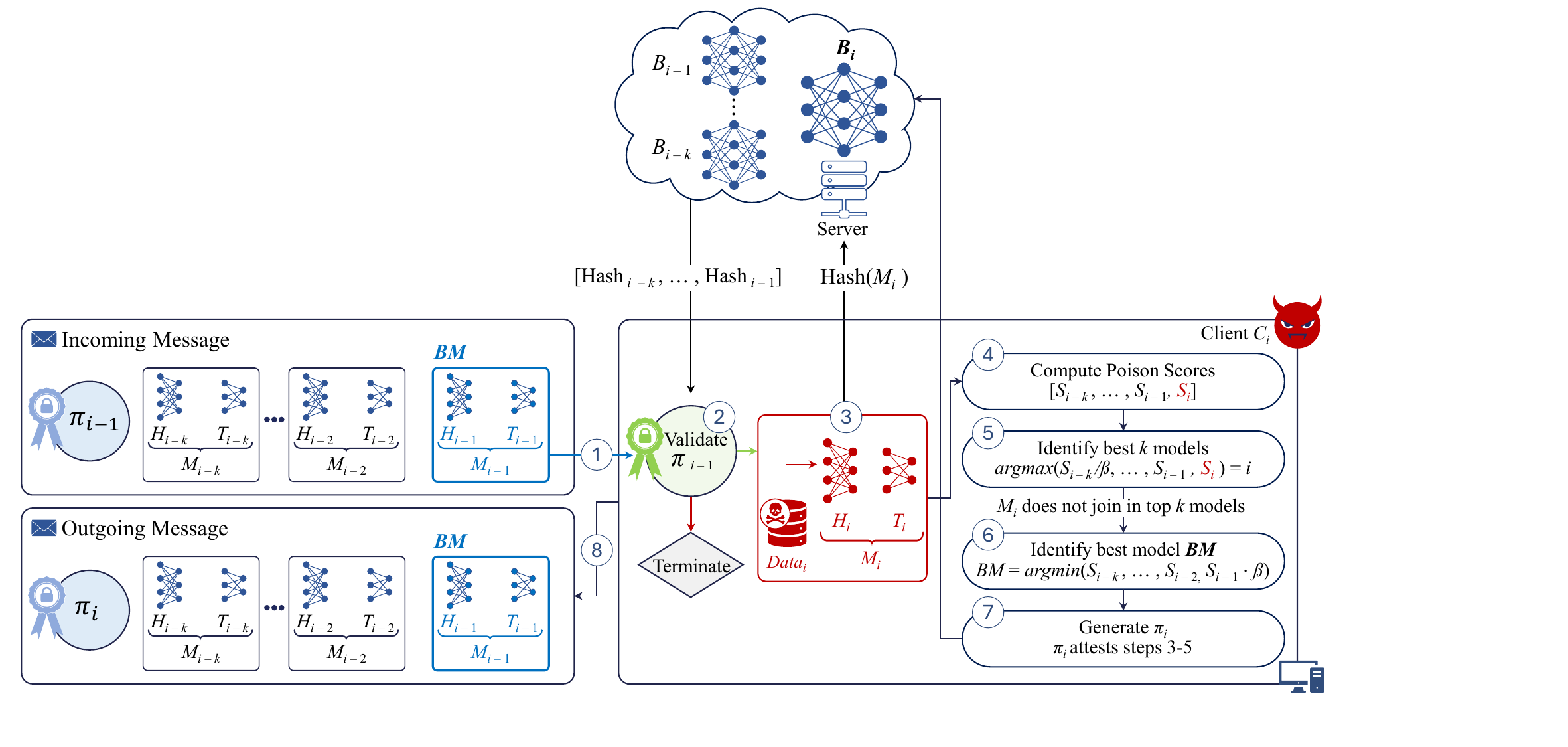}

    \caption{End-to-End Workflow of \ourname at Each Client in Split Learning.}
    \label{fig:highlevel}
\end{figure*}
\subsection{Requirements and Challenges}
An effective and practical defense against backdoor attacks needs to fulfill a number of requirements:\\
\textbf{R1 -- Mitigate Backdoor:} First, an effective defense needs to prevent \adversary from injecting the backdoor into the model.\\
\textbf{R2 -- Avoid Utility Degradation:} To be practical, the defense scheme must not negatively affect the model's utility, i.e., predictions for untriggered samples. \\
\textbf{R3 -- No model sharing of head and tail with server:} To ensure privacy and prevent that a party is enabled to analyze the parameters of the full model to violate clients' privacy, the defense must not share head or tail with the server.\\

\noindent A backdoor defense that effectively mitigates client-side backdoor attacks and fulfills these requirements faces a number of challenges:\\
\noindent\textbf{C1 – Detecting Poisoned Contributions Without Prior Knowledge.}  
A defining characteristic of backdoor attacks is their stealthiness. Given the large number of possible trigger patterns and target labels, defenders cannot make assumptions about the backdoor behavior, preventing a targeted inspection of the models. Furthermore, the data of different clients can differ from each other, be non-identical and independently distributed (non-IID). A major challenge, therefore, is distinguishing poisoned model updates from benign ones trained on non-IID data in the absence of knowledge about the backdoor.

\noindent\textbf{C2 – Partial Observability and Non-Comparable Models in Split Learning.}  
In SL, clients can only access specific partitions of the model (see R3), which limits their ability to perform comprehensive inspections. Additionally, due to the sequential nature of SL~\cite{rieger25safesplit}, each client trains its model update from a different checkpoint, resulting in models that are not directly comparable. A challenge is, therefore, how to reliably identify poisoned contributions when only partial model information is observable and the model updates are inherently \mbox{non-comparable due to sequential training.}

\noindent\textbf{C3 – Verifiable and Privacy-Preserving Client-Side Defense.}  
To offload computation from the server and enable scalable verification, we consider client-side defenses where each client analyzes its own model partition. However, malicious clients may manipulate this process to falsely validate poisoned updates. Also, requiring clients to inspect each other’s models introduces additional vulnerabilities, especially in the presence of colluding adversaries. A challenge is thus to enforce the correct execution of the defense mechanism and ensure that malicious behavior is exposed, while preserving client privacy and \mbox{maintaining robustness against collusion.}

\section{\sysNoSpace}

In the following, we outline the high-level design of \sys (\sect\ref{sec:sys-highlevel}) and \ournameGen flow (\sect\ref{sec:sys-flow}), before elaborating on how it inspects the client-side models for poisoned artifacts (\sect\ref{sec:sys-frequency}), how \ourname employs interactive ZKPs to guarantee the defense correctness (\sect\ref{sec:sys-zkp}) and discuss \ournameGen security parameters (\sect\ref{sec:sys-parameters}). In \sect\ref{sec:sys-implementation}, we describe the details of the privacy-preserving and efficient end-to-end implementation of \sys. The high-level approach towards securing SL training with \sys is shown in Fig.~\ref{fig:highlevel}.

\subsection{High-Level Overview}
\label{sec:sys-highlevel}

\ourname ensures the integrity of the collaborative training process through a client-side model inspection. The ZKP enforces that clients inspect their own model for backdoor artifacts in the \changedEditorial{frequency domain} and malicious clients expose themselves.

When it is client $C_i$'s turn, it receives a set of \queueLength model checkpoints, along with a pointer \bestModel indicating the most reliable model to continue training from. The client uses the model \bestModel to perform local training via standard U-shaped Split Learning and appends the resulting client-side model partition to the list, resulting in a total of $\queueLength + 1$ models. To assess the integrity of each model, a poisoning detection function is applied, assigning a risk score to every model in the list. The model with the highest poisoning score is subsequently removed, and \bestModel is updated to reference the most trustworthy remaining model. To ensure verifiable compliance with the defense mechanism, the client generates an interactive zero-knowledge proof $\pi$, which attests to the correct application of the protocol. This proof, together with the updated pointer and model list, is then forwarded to the next client.

A key aspect \changedEditorial{of \ournameGen design} is that the employed zero-knowledge proofs (ZKPs) are both sound and complete, meaning that a client cannot falsely attest to the correct execution of the protocol and still produce a valid proof. Consequently, any failed ZKP can be unambiguously traced back to the responsible client, allowing both the server and other clients to identify the malicious actor with certainty. This forces adversarial clients to strictly follow the \ourname protocol and execute the defense procedure on their own (potentially poisoned) model. This enforcement mechanism introduces a strategic dilemma for the adversary. On one hand, deviating from the protocol or manipulating the ZKP will result in the client being exposed and excluded. On the other hand, correctly executing the defense mechanism on a poisoned model will likely lead to its detection and removal. Since a malicious client’s goal is to have their poisoned model selected as the best model (\bestModel), they are incentivized to follow the protocol and ensure the generation of a valid proof, undermining the effectiveness of the attack itself.

\subsection{Defense Flow}
\label{sec:sys-flow}

The individual steps each client executes once the system is initialized are outlined below and illustrated in Fig.~\ref{fig:highlevel}.

\noindent\textbf{Step 1 -- Receive Model Checkpoints and Verification Data.}
At time step $t$, client $C_i$ receives from the previous client $C_{i-1}$ a list of \queueLength model checkpoints\footnote{Each model $M_j$ consists of the respective head $H_j$ and tail $T_j$. Further, for notational simplicity, we denote the top-$k$ best models received by client $C_i$ as $[M_{i-k}, \ldots, M_i-1]$, which assumes that all $k$ clients that precede $C_i$ are benign. This is not always the case, but malicious clients will be detected and replaced by benign clients or dropped.} $M_{i-\queueLength}, \ldots, M_{i-1}$, their corresponding updates, a pointer \bestModel identifying the recommended model for continuation, and a zero-knowledge proof $\pi_{i-1}$ attesting to the correct execution of the defense mechanism. Simultaneously, client $C_i$ obtains from the server the corresponding model hashes $\text{Hash}_{i-\queueLength}, \ldots, \text{Hash}_{i-1}$ allow client $C_i$ verifying the integrity of the received models and updates.

\noindent\textbf{Step 2 -- Validate Proof and Integrity of Received Models.}
Client $C_i$ verifies the received zero-knowledge proof $\pi_{i-1}$ to confirm that the previous client correctly executed the defense mechanism. Additionally, it compares the model and update hashes against the values provided by the server to ensure the authenticity and integrity of the forwarded models (step \circled{2} in Fig.~\ref{fig:highlevel}).

\noindent\textbf{Step 3 -- Conduct Local Training.}
Based on the pointer \bestModel, client $C_i$ instructs the server to use the corresponding backbone $B_i$ from the server’s stored list $B_{i-\queueLength}, \ldots, B_{i-1}$. $C_i$ then trains a new model $M_i$ using its private dataset and the referenced model checkpoint, following the standard U-shaped \changedEditorial{SL} protocol (step \circled{3} in Fig.~\ref{fig:highlevel}).

\noindent\textbf{Step 4 -- Poison Risk Scoring.}
A poison-scoring function is applied to all $\queueLength + 1$ models to determine a risk score that each model was trained on poisoned data. The scoring mechanism is based on the observation that backdoor attacks aim to change the predictions for triggered inputs toward the backdoor target label that differs from the correct label, thus being in contradiction to the model's benign behavior. This behavioral conflict introduces detectable artifacts in the frequency representation of the model updates. The scoring function captures the magnitude of these artifacts and assigns a corresponding risk score to each model (step \circled{4} in Fig.~\ref{fig:highlevel}).

\noindent\textbf{Step 5 -- Prune Model List.}
Using the computed poison-risk scores, one model is selected for removal from the list that $C_i$ is going to send to the next client $C_{i+1}$. This step serves a dual purpose. It eliminates potentially poisoned models that could compromise subsequent clients, particularly in the case of a malicious $C_i$. Further, when no poisoned model is detected, it ensures training progress by removing the oldest model. To encourage the removal of the oldest model in benign scenarios, its score is adjusted by dividing it by a security parameter $0 < \betaParam \leq 1$, increasing its relative score (step \circled{5} in Fig.~\ref{fig:highlevel}).

\noindent\textbf{Step 6 -- Update Best Model Pointer \bestModel.}
Following model pruning, the pointer \bestModel is updated to reference the model with the lowest poison risk score, which will be used by the next client $C_{i+1}$ as a checkpoint for the local training. To encourage the use of the most recent model, when it is not poisoned, and to support continued learning progress, the score of the newest model is scaled by \betaParam, reducing its score when it is likely benign.

\noindent\textbf{Step 7 -- Correctness Verification.} Client $C_i$ then interacts with the server and the next client $C_{i+1}$ in parallel to provide an interactive ZKP $\pi_i$ that attests to the correct calculation of poison scores (step 4), the worst model (step 5) and the other top-$k$ models (step 5), as well as \bestModel (Step 6). Additionally, client $C_i$ submits the hash $\text{Hash}_i$ of its newly trained model $M_i$ as well as the hash of the update $M_i - M_{i-1}$ to the server. The proof $\pi_i$ also verifies that the top-$k$ models were derived using only models committed to via the server-provided hashes $\text{Hash}_{i-k}, \ldots, \text{Hash}_{i-1}$ and that the update is actually the difference between $M_i$ and $M_{i-1}$ (step \circled{7} in Fig.~\ref{fig:highlevel}).

\noindent\textbf{Step 8 – Forward Models and Metadata.}
In the final step, client $C_i$ sends the updated list of the \queueLength best models and their corresponding updates, as determined in Step 5, along with the updated pointer \bestModel, to the next client $C_{i+1}$ (step \circled{8} in Fig.~\ref{fig:highlevel}).

\subsection{Poisoning Detection}
\label{sec:sys-frequency}

To detect poisoned training contributions, \ourname analyzes the updates to the head and tail model partitions in the frequency domain. The rationale is that, during the early stages of training, the introduction of new behaviors, such as those caused by backdoor attacks, leads to significant shifts in the \textit{low-frequency components} of the model's spectral representation~\cite{xu2019training,Rahaman2019spectralbias,rieger25safesplit}. Since the backdoor target label, by definition, differs from the benign label, the backdoor behavior contradicts previously learned benign behavior. Thus, the model must revert from benign to backdoor behavior, resulting in further detectable modifications in these low-frequency components. This means that the presence of new and conflicting behaviors introduced during local training can be effectively captured by measuring deviations in the model update's low-frequency spectrum.
\begin{algorithm}[tb]
    \caption{Poisoning Detection in \ourname at Step $i$}
    \begin{center}
        \scalebox{0.8}{
            \begin{minipage}{1.2\linewidth} 
                \begin{algorithmic}[1]
\Input 
    Model list $\mathbb{M}_{i-1} = [M_{i-\queueLength}, \ldots, M_{i-1}]$ \\
    Update list $\mathbb{U}_{i-1} = [U_{i-\queueLength}, \ldots, U_{i-1}]$ \\
    New model $M_i$, new update $U_i$, security parameter $\betaParam \in (0,1]$
\Output 
    Updated model list $\mathbb{M}_i$, update list $\mathbb{U}_i$, best model index $\text{BM}$

\Function{DetectPoisoned}{$\mathbb{M}_{i-1}, \mathbb{U}_{i-1}, M_i, U_i, \betaParam$}
    \State $\mathbb{M}_{\text{tmp}} \gets \mathbb{M}_{i-1} \cup \{M_i\}$
    \State $\mathbb{U}_{\text{tmp}} \gets \mathbb{U}_{i-1} \cup \{U_i\}$
    \State $\mathcal{S} \gets [\ ]$ \Comment{Initialize poison score list}

    \For{each $U_t \in \mathbb{U}_{\text{tmp}}$}
        \State \changedEditorial{$L_t \gets \text{LowFreq}(\text{DCT}(U_t))$}
        \State $s_t \gets \text{Taxicab}(L_t)$
        \State $\mathcal{S} \leftarrow \text{append}(\mathcal{S}, s_t)$
    \EndFor

    \State $\mathcal{S}[0] \gets \mathcal{S}[0] / \betaParam$
    \State $\mathcal{S}[\queueLength] \gets \mathcal{S}[\queueLength] \cdot \betaParam$

    \State $j \gets \text{\textit{argmax}}(\mathcal{S})$

    \State $\mathbb{M}_i \gets [M_t \in \mathbb{M}_{\text{tmp}} \mid t \neq j]$
    \State $\mathbb{U}_i \gets [U_t \in \mathbb{U}_{\text{tmp}} \mid t \neq j]$

    \State $\text{BM} \gets \text{\textit{argmin}}(\mathcal{S} \setminus \{\mathcal{S}[j]\})$
    \State \Return $\mathbb{M}_i, \mathbb{U}_i, \text{BM}$
\EndFunction
\end{algorithmic}

            \end{minipage}
        }
    \end{center}
\end{algorithm}

To compute the poison risk scores, \ourname first calculates the frequency-domain representation of each of the $\queueLength + 1$ model updates. It then isolates the relevant low-frequency components (see App.~\ref{app:low-frequencies}) and applies the \textit{Taxicab norm} (also known as $\ell_1$-norm) to quantify deviations. This choice of norm ensures robustness against outliers and emphasizes aggregate discrepancies across the spectrum.

To ensure training progress in the absence of poisoned models, \ourname adjusts the scores of specific models using a tunable \textit{security parameter} $\betaParam \in (0,1]$. Specifically, the score of the \textit{oldest model} is divided by $\betaParam$, increasing its relative likelihood of removal, while the score of the \textit{most recent model}, which was trained by the current client, is scaled by $\betaParam$ to slightly incentivize the selection of the most up-to-date benign model as the checkpoint for \changedEditorial{subsequent rounds.}

\subsection{Initialization}
\label{sec:sys-init}
\changedRevision{1}{The poisoning detection in \ourname relies on comparing poison risk scores across different models and selecting the one with the lowest score to update \bestModel. This mechanism requires initialization with at least one benign model to ensure a benign alternative reference is available, especially for cases where the first clients in the training queue are all malicious. Several strategies exist for this initialization. The most convenient approach is to use a (small) clean dataset or a pretrained model, if available. We introduce a secure initialization phase in which we temporarily increase the size of the reference list from $\queueLength$ to \numberOfClients. Each client independently trains its local model for one round, all starting from the same random model provided by the server. The resulting models are added to the reference list, and before the first effective round of Split Learning, the last client applies \ourname to select the \bestModel from this expanded reference set. As shown in Figure~\ref{fig:eval-init}, this initialization process is agnostic to client ordering, consistently identifies a benign starting model, and achieves performance comparable to using a pre-trained model. As such, the secure initialization phase can be seamlessly interchanged with a pre-trained model when one is available, making both approaches functionally equivalent in terms of effectiveness and resulting model integrity.}

\subsection{Cryptographic Attestation of \changedEditorial{\ourname}}
\label{sec:sys-zkp}

\begin{algorithm}[tbp] 
    \caption{\sys game} 
    \label{alg:zkp_resized}
    \noindent 
    
    \resizebox{\columnwidth}{!}{
    \begin{tabular}{
    @{} >{\raggedright\arraybackslash}p{0.42\columnwidth} 
    @{\hspace{1em}} c @{\hspace{1em}} 
    @{} >{\raggedright\arraybackslash}p{0.42\columnwidth} 
    @{}
}
    \multicolumn{3}{@{}l@{}}{\textbf{Public Inputs (\Prv \& \Vrf):} $[\text{Hash}_{i-k},\cdots,\text{Hash}_{i}]$, $[\text{Hash}_{DCT(M_{i-k}}),\cdots,\text{Hash}_{DCT(M_{i})}]$, $S_{BM}$, $S_{WM}$} \\
    \multicolumn{3}{@{}l@{}}{\textbf{Private Inputs (\Prv):} [$M_{i-k}, \cdots, M_{i}$], [$DCT(M_{i-k}), \cdots, DCT(M_i)$]} \\
    \hline \\[-1.5ex] 
    \multicolumn{1}{@{}l}{\textbf{Prover \Prv (Client $C_i)$}} & & \multicolumn{1}{l@{}}{\textbf{Verifier \Vrf (Client $C_{i+1}$ \& Server)}} \\
    \hline \\[-1.5ex] 


    \textbf{ZKP} $\pi \leftarrow$ \textbf{ZK Circuit:}  & &\\
    \textbf{begin circuit} & & \\
    \multicolumn{2}{@{}l}{\textbf{for} $M_t$ \textbf{in} [$M_{i-k}, \cdots, M_{i}$]:} & \\
    \multicolumn{2}{@{}l}{\ \ \ \ \ \textbf{assert} Hash($M_t$) = Hash$_{t}$} & \\
    \multicolumn{2}{@{}l}{\ \ \ \ \ \textbf{compute} $S_t$ = $S(M_t)$} & Compute randomness $r$ \\
    \ \ \ \ \ Receive $r$ & \hspace{0.5em} $\xleftarrow{\hspace*{1.5cm} r \hspace*{1.5cm}}$ \hspace{0.5em} & Send $r$ \\ 
    
    \multicolumn{2}{@{}l}{\ \ \ \ \ $\hookrightarrow$ \textbf{assert } $\text{Hash}(DCT(M_t))=\text{Hash}_{DCT(M_t)}$} & \\
    \multicolumn{2}{@{}l}{\ \ \ \ \ $\hookrightarrow$ \textbf{assert } $DCT(M_t,r)$ = $DCT(M_t)\cdot r$} & \\
    \textbf{end for}\\    
    \multicolumn{2}{@{}l}{\textbf{assert} $S_{WM}$ = max($S_{i-k} / \beta, \cdots, S_i$)} & \\
    \multicolumn{2}{@{}l}{\textbf{assert} $S_{BM}$ = min($S_{i-k}, \cdots, S_i \cdot \beta$)} & \\
    \textbf{end circuit}
    \\[0.6ex] 

    Send $\pi$ & \hspace{0.5em} $\xrightarrow{\hspace*{1.5cm} \pi \hspace*{1.5cm}}$ \hspace{0.5em} & Receive $\pi$ \\ 

    & & Verify proof $\pi$: \\
    & & \textbf{Output:} Accept / Reject \\ 

    \hline 
    \multicolumn{3}{@{}l@{}}{\textbf{Note:} \Prv only wins game if $\pi$ is accepted by \Vrf. $\pi$ is only accepted if all operations in}\\
    \multicolumn{3}{@{}l@{}}{\textbf{ZK Circuit} are computed soundly and with valid inputs.}\\
\end{tabular}%
    } 
\end{algorithm}

Due to the client-based nature of \sys, malicious clients may stray from the protocol in an attempt to adversely affect the training process without being detected. Although \sys introduces an approach to dropping poisoned models based on their poison risk score, if the integrity of the poison risk score calculations is not ensured, then malicious clients can still collude and remain undetected. In Alg.~\ref{alg:zkp_resized}, we outline \ournameGen approach to probabilistically guarantee faithful execution of the proposed client-side defense scheme while ensuring that client privacy is still maintained through the use of interactive ZKPs. 

In ZK terminology, a "circuit" is defined as any arbitrary computation that a prover aims to attest the computational integrity and compliance of. Alg.~\ref{alg:zkp_resized} defines a circuit that is evaluated by a prover \Prv, resulting in a ZKP $\pi$. This circuit is evaluated on public and private inputs, in which public inputs are known to both the verifier \Vrf and prover \Prv, and the private inputs are only known to the prover \Prv. The public inputs are the committed values of the top-$k$ models and client $C_i's$ \changedEditorial{model}, alongside the committed values of these models after undergoing the discrete cosine transform (DCT). Poison risk scores corresponding to the best (smallest poison risk) and worst (highest poison risk) $S_{BM}$ and $S_{WM}$, respectively, are also public inputs. The only private inputs are the unmasked, plaintext model parameters of the top-$k$ models and client $C_i$'s model alongside their frequency representations. 
In the proposed setting, the prover \Prv is a client $C_i$ who has already trained their model $M_i$ on their private dataset $D_i$ on the backbone $B_i$ hosted on the server. The goal of the prover is to concurrently convince client $C_{i+1}$ and the server, who each act as a verifier \Vrf, that they are faithfully executing the required computations for the \sys defense scheme. The prover shares $\pi$ with both the following client $C_{i+1}$ \textit{and} the server to ensure that if client $C_{i+1}$ is malicious, then it cannot falsely reject the proof.

The initial loop iterates through each model to ensure correct analysis. For each model $M_t$, the first assertion $\text{Hash}(M_t) = \text{Hash}_t$ guarantees that the prover is indeed using the expected model corresponding to a previously published, publicly known hash. This prevents the prover from substituting a different, potentially benign-looking model just to generate a valid proof. Subsequently, the poison risk score $S_t$ is computed based on the authenticated model $M_t$. One of the core operations that enables an accurate poison risk score is the DCT, which is a costly operation to verify in ZK. To address this, we take a modified approach using Freivald's algorithm \cite{freivalds1979fast}, which utilizes verifier-generated randomness $r$ to verify the correct calculation of the DCT and compares the committed result to a publicly known hash of the result to convince the verifier of the computational integrity. The circuit takes in the DCT of each model as a private input to allow for further consistency checks. This will be discussed in detail in section \ref{sec:sys-implementation}. This sequence tightly binds the original model and its calculated scores to the publicly available commitments, taken in as public inputs to the ZK circuit, to ensure faithful execution of the operations on each model.

Following the per-model checks, the circuit performs crucial aggregate verifications using the final two assertions. These checks concern the claimed best model score $S_{BM}$ and the worst model score $S_{WM}$ within the relevant window (models $i-k$ through $i$, including the prover $C_i$'s own model $M_i$). These assertions ensure that the prover correctly identifies the worst model, in terms of poison risk score, and proves that it is not included in the top-$k$ models that it sends to the next client. Successfully passing these checks demonstrates that the prover not only computed the poison risk scores correctly, but also accurately modified the top-$k$ models to appropriately reflect the poison risk scores. If all assertions hold, the entire computation is soundly encoded into a proof $\pi_i$ that will be validated by the verifiers, the next client $C_{i+1}$ and the server. When the verifiers receive and successfully verify $\pi_i$, they gain high confidence that client $C_i$ followed the \sys protocol faithfully without releasing any sensitive information. A rejection of $\pi_i$ signals a deviation from the \sys protocol and potential malicious intent. This ZKP mechanism forms the trust anchor for the client-based \sys defense strategy.

\subsection{Security Parameters}
\label{sec:sys-parameters}

The poisoning detection mechanism of \ourname relies on two parameters, the \textit{security parameter} $\betaParam$ and the \textit{queue length} $\queueLength$. In the following, we elaborate on these parameters and their impact on the performance of \ourname. An ablation study evaluating different parameter configurations is provided \changedEditorial{in App.~\ref{app:ablation} to} complement the theoretical discussion provided here. 

\noindent\paragraph{Security Parameter \boldmath{$\betaParam$}.}  
The security parameter $\betaParam \in (0, 1]$ performs the trade-off between \ournameGen robustness against poisoned models and its ability to maintain a continuous training progress. The $\betaParam$ parameter is used to scale the risk score of the newest model, by multiplication, and the oldest model, by division. A smaller $\betaParam$ value reduces the used value of the most recent model's risk score, increasing the likelihood of it being selected for the next training iteration. Simultaneously, it increases the relative score of the oldest model, making it more likely to be pruned from the queue. This mechanism helps prioritize newer, potentially benign updates and avoid discarding recent training contributions unnecessarily. However, setting $\betaParam$ too low may cause the scheme to overly favor the most recent model, even in the presence of subtle poisoning, weakening the system’s security. We choose $\betaParam = 0.7$ to ensure a practical trade-off between training efficiency and defense robustness.

\noindent\paragraph{Queue Length \queueLength.}
The queue length \queueLength determines the number of past model checkpoints (and updates) that are retained and evaluated at each client during the training process. A larger \queueLength increases the defense’s robustness by providing a broader context for detecting poisoned updates and reducing the risk of coordinated attacks from malicious clients. On the other side, a lower number of shared client models reduces the privacy risk as fewer models are shared with each client for lower values, reducing the risk of successful privacy attacks. Further, smaller values reduce the communication and computational overhead, as each client must evaluate more models. In our experiments in \sect\ref{sec:eval}, we choose \queueLength=3.

\subsection{Implementation}
\label{sec:sys-implementation}

\noindent \textbf{ZKP Circuit Implementation. }
\ourname utilizes Zero-Knowledge Proofs (ZKPs) based on Vector Oblivious Linear Evaluation (VOLE), leveraging its advantageous scalability compared to alternative ZKP schemes, a key finding highlighted in \cite{sheybani2025zero}. As discussed in \sect\ref{sec:prelims-zkp}, many other ZKP constructions involve computationally demanding setup phases, which might necessitate significant communication overhead or reliance on a trusted third party. VOLE-based ZK strikes an effective balance between computational efficiency during proof generation and verification and scalability, making it well-suited for our application.

The ZK circuit, detailed in Alg.~\ref{alg:zkp_resized}, is realized using the \texttt{emp-zk} library \cite{empzk}, a state-of-the-art framework specifically designed for VOLE-based ZK protocols. \texttt{emp-zk} is built upon the \texttt{emp-tool} toolkit \cite{emptool}, a widely adopted library for efficient multi-party computation. To optimize performance, we employ a customized version of the \texttt{emp-tool} backend, specifically enhanced to support parallel proof generation and further multithreading capabilities, reducing latency of proof generation and verification.

Our ZK circuit design accommodates computations involving both arithmetic and Boolean values. Arithmetic circuits are employed for operations like addition and multiplication, primarily used when computing the poison risk scores ($S_t$). Boolean circuits are utilized for operations such as cryptographic hashing. The seamless integration of both circuit types is facilitated by the efficient conversion techniques presented in \cite{weng2021mystique}, enabling a fine-grained optimization strategy for complex computations. Specifically, commitments within the ZK circuit are computed and verified using a custom Boolean SHA-256 implementation. This implementation efficiently processes model parameters, represented as serialized bitstreams, to produce a 256-bit cryptographic hash digest.

Instead of a monolithic design, we adopt a modular approach to building the ZK circuits in \ourname. This design choice promotes extensibility, enabling the reuse of core components (e.g., DCT, SHA-256, matrix multiplication) across different client-side defense schemes. Developers can also integrate new defense-specific modules, such as $\ell_2$-norm or thresholding, within the same framework.

\noindent \textbf{Speeding Up DCT Verification.} 
The core operation influencing the poison risk score calculation is the Discrete Cosine Transform (DCT) applied to model parameters $M_t$, represented as a matrix of size $\left\lceil\sqrt{|H_t + T_t|}\right\rceil\times\left\lceil\sqrt{|H_t + T_t|}\right\rceil$, where $H_t$ is the number of head parameters and $T_t$ is the number of tail parameters. Computing this matrix directly within a ZK circuit is highly inefficient due to the required cosine functions and the large number of multiplications and additions over potentially very large matrices. The conversions between arithmetic and Boolean domains for cosine approximations would introduce significant overhead.

To mitigate this complexity, we avoid computing the DCT directly within the ZK circuit. Instead, we leverage a technique inspired by Freivald's algorithm, a classical method for probabilistically verifying matrix multiplication \cite{freivalds1979fast}. Through our experiments, we found that this approach was around 2 orders of magnitude faster than the naive approach of computing the DCT within the ZK circuit. Freivald's algorithm allows checking if $C = A B$ for matrices $A, B, C$ by choosing a random binary vector $r$ and verifying if $C r = A (B r)$. This check is significantly faster than performing the full matrix multiplication $A B$. If $C = A B$, the equality $C r = A (B r)$ always holds. If $C \neq A B$, the equality holds only with a small probability. This process can be repeated many times to ensure high probabilistic guarantees.
\begin{equation} \label{eq:freivalds}
\text{Check: } C r \stackrel{?}{=} A (B r) \quad \text{for random } r
\end{equation}
This probabilistic verification fits naturally within interactive ZK protocols, where \Vrf can provide the random challenge vector $r$. \Prv then performs the computations involving $r$ inside the ZK circuit to convince the Verifier of the original matrix relationship's validity without revealing the private matrices (like $M_t$ in our case).

Applying this idea to our DCT scenario, the 2D DCT can be expressed in matrix form as $D_t = C_{N} M_t (C_N)^T$, where $C_N$ represents a DCT transformation matrix, respectively, whose entries depend on the cosine terms from the standard definition of the DCT. This is how the DCT is efficiently computed on the \texttt{opencv} C++ package \cite{OpenCV}. The matrix $C_N$ is known, fixed and depends only on the dimension $\left\lceil\sqrt{|H_t + T_t|}\right\rceil$. It can be pre-computed offline and treated as a constant within the ZK circuit. \Prv's goal is not to explicitly re-compute $DCT(M_t)$ inside the circuit but rather to use a randomized check to prove that the privately held $DCT(M_t)$ is consistent with the privately held $M_t$, thus ensuring the integrity of the inputs for the calculation of the poison risk score $S_t$.

We achieve this using the randomized check provided by \Vrf's random vector $r$ (of dimension $T_t \times 1$). Instead of checking the full matrix equality $DCT(M_t) = C_{N} \cdot M_t \cdot (C_{N})^T$, we check the related equality probabilistically by multiplying by $r$: $DCT(M_t) \cdot r = (C_{N} \cdot M_t \cdot (C_{N})^T) \cdot r$. The right-hand side can be computed efficiently within the arithmetic ZK circuit using matrix-vector multiplications by exploiting associativity:
\begin{enumerate}
    \item Compute $v_1 = (C_N)^T \cdot r$. (Size $\left\lceil\sqrt{|H_t + T_t|}\right\rceil \times 1$)
    \item Compute $v_2 = M_t \cdot v_1$. (Size $\left\lceil\sqrt{|H_t + T_t|}\right\rceil \times 1$)
    \item Compute $Z = C_N \cdot v_2$. (Size $\left\lceil\sqrt{|H_t + T_t|}\right\rceil \times 1$)
\end{enumerate}
This computation, denoted as $\text{DCT}(M_t, r)$ in Alg.~\ref{alg:zkp_resized}, involves only matrix-vector products using \Prv's private matrix $M_t$, the constant cosine transformation matrix $C_N$, and \Vrf's randomness $r$. All operations are arithmetic (multiplications and additions), which are efficiently handled by \texttt{emp-zk}'s arithmetic circuit backend. The final step within the ZK circuit is to assert that $Z$ matches the corresponding private input times the random vector $r$, $DCT(M_t)\cdot r$.
This ensures, with high probability, that \Prv is correctly computing the poison risk score $S_t$ for model $M_t$.

\section{Experimental Evaluation}
\label{sec:eval}

\begin{table}[t]
    \caption{Overview of default experiment parameters.}
    \label{tab:eval-setup:defaultparams}
    \centering
    \scaleTable{\begin{tabular}{l|l}
\textbf{Hyperparameter} & \textbf{Value} \\
\hline
\changedRevision{6}{Poison Data Rate (PDR)} & 75\% \\
Number of Clients & 10\\
\changedRevision{6}{Poison Model Rate (PMR)} & \changed{20\%}\\
IID Degree& 0.8\\
Dataset&\cifar\\
DNN Architecture& ResNet18\\
\betaParam&0.7\\
Queue Length \queueLength & 3\\
Initialization & Pretrained\\
\end{tabular}}
\end{table}

In the following, we perform an extensive evaluation of \ournameGen \\ \changed{robustness} using various settings and also measure the runtime overhead of the ZKP-protocol. Further, in App.~\ref{app:ablation}  we provide an ablation study.

\subsection{Experimental Metrics}
\label{sec:eval-metrics}

In our experiments, we evaluate the effectiveness \changedEditorial{of \ourname on the model} using four metrics: \textit{Backdoor Accuracy (BA)}, \textit{Main Task Accuracy (MA)}, \textit{Poisoned Removal Rate (PRR)}, and \textit{Better Benign Rate (\changedEditorial{BBR})}. These metrics capture both security-related and utility aspects of the training process.

\noindent\textbf{Backdoor Accuracy (BA)} measures the effectiveness of the backdoor embedded in a model. It is computed as the model’s accuracy on a backdoored test set, i.e., a set of inputs containing the trigger pattern. The BA indicates the proportion of triggered samples classified as the attacker’s target label. The adversary $\adversary$ seeks to maximize BA, whereas an effective defense aims to minimize it (see R1 in \sect\ref{sec:problem-requirements}).

\noindent\textbf{Main Task Accuracy (MA)} reflects the model's utility on the intended benign task. It is evaluated as standard accuracy on an unaltered, benign test set, following non-security-focused deep learning benchmarks. Both the adversary $\adversary$ and the defense seek to preserve a high MA, the \adversary to remain undetected, and the defense to ensure practical applicability (see requirement R2 in \sect\ref{sec:problem-requirements}).

\noindent\textbf{Poisoned Removal Rate (PRR)} quantifies the defense's effectiveness in eliminating poisoned models. It is defined as the fraction of time steps in which at least one poisoned model was present in the queue and the model selected for removal was indeed poisoned. This metric only considers scenarios where poisoned models are present and removable. To prevent any backdoor from persisting, an effective defense must strive to maximize PRR by promptly removing newly introduced poisoned models.

\noindent\textbf{Better Benign Rate (\changedEditorial{BBR})} captures the frequency with which the defense fails to select the most recent benign model despite its availability. A high \changedEditorial{BBR} suggests that the defense unnecessarily reverts to older models, potentially hindering convergence and training progress. Therefore, a robust defense should aim to minimize \changedEditorial{BBR}, promoting continuous learning while ensuring model integrity.

\subsection{Experimental Setup}
\label{sec:eval-setup}

\begin{table}[t]    
    \caption{Effectiveness of \ourname for different datasets in terms of Backdoor Accuracy (BA), Main Task Accuracy (MA), Poisoned Removal Rate (PRR), and Better Benign Rate (BBR).}
    \label{tab:eval-results:datasets}
    \centering
    \scaleTable{
\begin{tabular}{ll|cccc}
	Dataset & Defense & BA & MA & PRR & BBR\\\hline
	\multirow{4}{*}{Cifar10} & Benign & 3.71\% & 74.63\% & \multicolumn{1}{c}{-} & \multicolumn{1}{c}{-} \\
	& No Defense & 93.29\% & 70.56\% & \multicolumn{1}{c}{-} & \multicolumn{1}{c}{-} \\
	& \ourname & 4.45\% & 73.02\% & 86.10\% & 8.02\% \\
	& Gold Standard & 3.30\% & 73.40\% & 100.00\% & 0.00\% \\
	\hline
	\multirow{4}{*}{MNIST} & Benign & 0.23\% & 99.29\% & \multicolumn{1}{c}{-} & \multicolumn{1}{c}{-} \\
	& No Defense & 100.00\% & 67.72\% & \multicolumn{1}{c}{-} & \multicolumn{1}{c}{-} \\
	& \ourname & 36.99\% & 99.25\% & 100.00\% & 3.79\% \\
	& Gold Standard & 0.00\% & 99.24\% & 100.00\% & 0.00\% \\
	\hline
	\multirow{4}{*}{FMNIST} & Benign & 1.19\% & 86.65\% & \multicolumn{1}{c}{-} & \multicolumn{1}{c}{-} \\
	& No Defense & 100.00\% & 84.67\% & \multicolumn{1}{c}{-} & \multicolumn{1}{c}{-} \\
	& \ourname & 1.38\% & 86.22\% & 58.81\% & 6.57\% \\
	& Gold Standard & 1.46\% & 86.40\% & 100.00\% & 0.00\% \\
	\hline
    
    
    \multirow{4}{*}{\changedRevision{5}{\forest}} & \changed{Benign} & \changed{0.00\%} & \changed{99.79\%} & \multicolumn{1}{c}{\changed{-}} & \multicolumn{1}{c}{\changed{-}} \\
	& \changed{No Defense} & \changed{87.40\%} & \changed{45.37\%} & \multicolumn{1}{c}{\changed{-}} & \multicolumn{1}{c}{\changed{-}} \\
	& \changed{\ourname} & \changed{0.03\%} & \changed{99.77\%} & \changed{94.99\%} & \changed{14.36\%} \\
	& \changed{\goldenDefense} & \changed{0.00\%} & \changed{99.83\%} & \changed{100.00\%} & \changed{0.00\%} \\
	\hline
	\multirow{4}{*}{\changedRevision{5}{\cifarhundred}} & \changed{Benign} & \changed{0.24\%} & \changed{57.11\%} & \multicolumn{1}{c}{\changed{-}} & \multicolumn{1}{c}{\changed{-}} \\
	& \changed{No Defense} & \changed{99.98\%} & \changed{47.04\%} & \multicolumn{1}{c}{\changed{-}} & \multicolumn{1}{c}{\changed{-}} \\
	& \changed{\ourname} & \changed{0.30\%} & \changed{55.86\%} & \changed{100.00\%} & \changed{15.79\%} \\
	& \changed{\goldenDefense} & \changed{0.25\%} & \changed{56.27\%} & \changed{100.00\%} & \changed{0.00\%} \\
	\hline
	\multirow{4}{*}{\changedRevision{5}{\tinyimagenet}} & \changed{Benign} & \changed{0.09\%} & \changed{66.36\%} & \multicolumn{1}{c}{\changed{-}} & \multicolumn{1}{c}{\changed{-}} \\
	& \changed{No Defense} & \changed{99.96\%} & \changed{58.50\%} & \multicolumn{1}{c}{\changed{-}} & \multicolumn{1}{c}{\changed{-}} \\
	& \changed{\ourname} & \changed{0.07\%} & \changed{66.90\%} & \changed{94.10\%} & \changed{10.19\%} \\
	& \changed{\goldenDefense} & \changed{0.10\%} & \changed{66.74\%} & \changed{100.00\%} & \changed{0.00\%} \\
    
\end{tabular}}
\end{table}

\begin{table}[bt]
    \caption{Effectiveness of \ourname for different model architectures in terms of Backdoor Accuracy (BA), Main Task Accuracy (MA), Poisoned Removal Rate (PRR), and Better Benign Rate (BBR).}
    \label{tab:eval-results:models}
    \centering
    \scaleTable{\begin{tabular}{ll|cccc}
	Model & Defense & BA & MA & PRR & BBR\\\hline
	\multirow{4}{*}{GoogLeNet~\cite{szegedy2015going}} & Benign & 4.20\% & 67.14\% & \multicolumn{1}{c}{-} & \multicolumn{1}{c}{-} \\
	& No Defense & 98.61\% & 59.85\% & \multicolumn{1}{c}{-} & \multicolumn{1}{c}{-} \\
	& Gold Standard & 3.85\% & 65.65\% & 100.00\% & 0.00\% \\
	& \ourname & 5.23\% & 65.35\% & 80.09\% & 4.12\% \\
	\hline
	\multirow{4}{*}{MicronNet~\cite{wong2018micronnet}} & Benign & 4.23\% & 68.85\% & \multicolumn{1}{c}{-} & \multicolumn{1}{c}{-} \\
	& No Defense & 99.51\% & 53.55\% & \multicolumn{1}{c}{-} & \multicolumn{1}{c}{-} \\
	& Gold Standard & 5.78\% & 67.97\% & 100.00\% & 0.00\% \\
	& \ourname & 5.64\% & 68.01\% & 55.48\% & 3.12\% \\
	\hline
	\multirow{4}{*}{ResNet-18~\cite{he2016deep}} & Benign & 3.71\% & 74.63\% & \multicolumn{1}{c}{-} & \multicolumn{1}{c}{-} \\
	& No Defense & 93.29\% & 70.56\% & \multicolumn{1}{c}{-} & \multicolumn{1}{c}{-} \\
	& Gold Standard & 3.30\% & 73.40\% & 100.00\% & 0.00\% \\
	& \ourname & 4.45\% & 73.02\% & 86.10\% & 8.02\% \\
	\hline
	\multirow{4}{*}{ResNet34~\cite{he2016deep}} & Benign & 2.18\% & 69.74\% & \multicolumn{1}{c}{-} & \multicolumn{1}{c}{-} \\
	& No Defense & 100.00\% & 66.24\% & \multicolumn{1}{c}{-} & \multicolumn{1}{c}{-} \\
	& Gold Standard & 2.15\% & 69.20\% & 100.00\% & 0.00\% \\
	& \ourname & 2.69\% & 69.17\% & 85.98\% & 3.12\% \\
	\hline
	\multirow{4}{*}{VGG11~\cite{simonyan2015very}} & Benign & 3.00\% & 69.55\% & \multicolumn{1}{c}{-} & \multicolumn{1}{c}{-} \\
	& No Defense & 99.98\% & 69.55\% & \multicolumn{1}{c}{-} & \multicolumn{1}{c}{-} \\
	& Gold Standard & 4.97\% & 69.07\% & 100.00\% & 0.00\% \\
	& \ourname & 4.82\% & 67.20\% & 55.48\% & 2.67\% \\
	\hline
	\multirow{4}{*}{WideResNet52~\cite{zagoruyko2016wide}} & Benign & 2.36\% & 80.84\% & \multicolumn{1}{c}{-} & \multicolumn{1}{c}{-} \\
	& No Defense & 99.97\% & 74.95\% & \multicolumn{1}{c}{-} & \multicolumn{1}{c}{-} \\
	& Gold Standard & 2.33\% & 80.30\% & 100.00\% & 0.00\% \\
	& \ourname & 2.15\% & 79.93\% & 100.00\% & 18.25\% \\
\end{tabular}}
\end{table}
\subsubsection{Training Setup}\label{sec:eval-setup-training}

\changedRevision{6}{In the following, we extensively investigate various settings for the training process. The default parameters of the experiment scenario are shown in Tab.~\ref{tab:eval-setup:defaultparams}, which are always used unless other parameters are described. Notably, in our default setting, we use 10 clients in total and a PMR of 20\%, resulting in 2 malicious clients. However, when increasing the client numbers in \sect\ref{sec:eval-setting} using the same PMR, the absolute number of malicious clients increases accordingly. The positions of the malicious clients are randomly chosen.} \changedRevision{5}{For the evaluation, we employ five image benchmark datasets (\cifar, \mnist, \fmnist, \cifarhundred, \tinyimagenet) that are typically used to assess the security of distributed learning~\cite{rieger25safesplit,tajalli2023feasibility,krauss2023mesas} and one tabular dataset, the \forest dataset. We describe them in App.~\ref{app:datasets}.}\\
\textbf{DNN Architectures.} While using ResNet-18 as the default architecture, we extend the evaluation to a wide range of other architectures to show \ournameGen scalability and generalizability. An overview of the leveraged models and the number of their parameters is shown in Tab.~\ref{tab:eval-setup:dnnparams}.\\

\changed{\textbf{Initialization.}} \changedRevision{1}{For our experiments, we use a pre-trained model for our default setting, obtained via 3 rounds of benign training in the respective setup. In \sect\ref{sec:eval-setting} we empirically analyze the impact of our initialization described in \sect\ref{sec:sys-init} on \ournameGen performance.}
\subsubsection{Computational Setup.}\label{sec:eval-setup-server} The ML experiments are conducted on two servers. One server is equipped with an AMD EPYC 9954, 576~GB of main memory, and 4 NVIDIA A6000. The second server, which was used for the experiments for different models (Tab.~\ref{tab:eval-results:models}) and datasets (Tab.~\ref{tab:eval-results:datasets}), as well as the ZKP runtime evaluations, was equipped with an AMD EPYC 7742, 4 NVIDIA Quadro RTX 8000 and 1 TB main memory. \changedRevision{4}{For measuring the main memory consumption (see \sect\ref{sec:eval-edge}) of the SL training, we used a Raspberry PI 4B with 4GB main memory.}\\
All experiments were conducted using Pytorch~\cite{pytorch}. While we ensured the reproducibility of the experiments by seeding the random generators, different library versions on both servers caused slightly varying results as the random generators were initialized differently. All experiments were repeated 3 times with different seeds, and the average results are shown. In App.~\ref{app:scattering}, we evaluate and discuss the variance in the results.

\begin{table}[t]
    \caption{Overview of employed Deep Neural Network (DNN) architectures.}
    \label{tab:eval-setup:dnnparams}
    \centering
    \scaleTable{\begin{tabular}{l|l|r|r|r}
 \textbf{Model} & \textbf{Dataset} & \multicolumn{3}{c}{\textbf{Number of Parameters}} \\
                  &                  & \textbf{Head} & \textbf{Backbone} & \textbf{Tail} \\\hline
    
        VGG11 & \cifar & \numprint{1920} & \numprint{13418880} & \numprint{5130} \\
        ResNet34 & \cifar & \numprint{9536} & \numprint{21275136} & \numprint{5130} \\
        GoogLeNet & \cifar & \numprint{9536} & \numprint{5590368} & \numprint{10250} \\
        WideResNet52 & \cifar & \numprint{9536} & \numprint{66824704} & \numprint{204900} \\
        ResNet-18 & \cifar & \numprint{9536} & \numprint{11166976} & \numprint{5130} \\
        ResNet-18 & \mnist & \numprint{3264} & \numprint{11166976} & \numprint{5130} \\
        ResNet-18 & \fmnist & \numprint{3264} & \numprint{11166976} & \numprint{5130} \\
        \changed{MicronNet} &  \changed{\cifar} & \changed{\numprint{824}} & \changed{\numprint{411192}} & \changed{\numprint{3010}} \\
        \changedRevision{5} {WideResNet52} &  \changed{\cifarhundred} &  \changed{\numprint{9536}} & \changed{\numprint{66824704}} & \changed{\numprint{204900}} \\
        \changedRevision{5} {WideResNet52} &  \changed{\tinyimagenet} & \changed{\numprint{9536}} & \changed{\numprint{66824704}} & \changed{\numprint{409800}}\\
        \changedRevision{5} {SimpleMLP} &  \changed{\forest} & \changed{\numprint{14848}} & \changed{\numprint{570336}} & \changed{\numprint{4357}} \\
        
\end{tabular}
}
\end{table}

\begin{figure}[t]
	\centering{	
			\includegraphics[width=0.775\columnwidth]{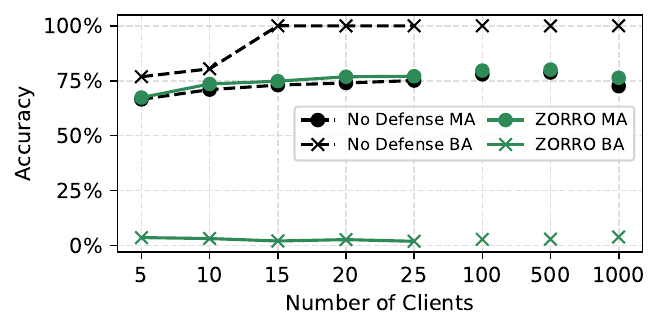}
	}
	\caption{\changedRevision{2}{\ournameGen effectiveness for different client numbers.}}
	\label{fig:eval-clients}
\end{figure}

\subsubsection{Baseline Defenses}
In our evaluation, we use two baselines, No Defense as well as a defense we denote as \goldenDefense. The second one is a hypothetical defense that knows the benign and poisoned models, allowing it to always exclude poisoned models and use the latest benign model for future training. While in practice, this defense is not possible, it shows the optimal performance that \ourname should achieve.

\subsection{High-Level Results}
\label{sec:eval-overview}

Tab.~\ref{tab:eval-results:datasets} shows \ournameGen effectiveness for different datasets\footnote{\changedRevision{6}{Notably, \cifarhundred and \tinyimagenet deviate from the setting described in Tab.~\ref{tab:eval-setup:defaultparams}, as discussed in App.~\ref{app:datasets}.}}. As the table shows \ourname effectively mitigates the attack. \ourname effectively mitigates the attack for all datasets. For \cifar and \fmnist, it reduces the BA to a similar level as the \goldenDefense. Only for \mnist the BA is in average 36.99\%, although \ourname always excludes poisoned models, guaranteeing that the model is backdoor-free. 
The relatively high BA (36.99\%) observed on \mnist can be attributed to an artifact rather than a failure of the defense. Since \ourname excluded the poisoned models in every repetition, and only one out of three runs exhibited high BA, it is likely that the benign model was distracted by the trigger. In such cases, the model may consistently predict a specific label in the presence of the trigger, which, if it by chance is the backdoor target label, results in a high BA, even though the model is benign. We provide a detailed investigation of this phenomenon in Appendix~\ref{app:mnistAnomaly}.

Tab.~\ref{tab:eval-results:models} shows the effectiveness of \ourname across different model architectures. As the table shows, \ourname consistently mitigates the backdoor attacks, reducing BA to levels comparable with the \goldenDefense while maintaining high MA. Notably, although the PRR for MicronNet is only 55.48\%, the BA remains low at 5.64\%, showing that the poisoned models were either neutralized or ineffective. Similarly, for WideResNet52, \ourname achieves perfect PRR (100.00\%) and the lowest BA (2.15\%) among all defenses, showcasing its robustness even for large-scale models. Overall, \ourname is effective across a diverse range of architectures.

\begin{figure}[t]
	\centering{	
			\includegraphics[width=0.775\columnwidth]{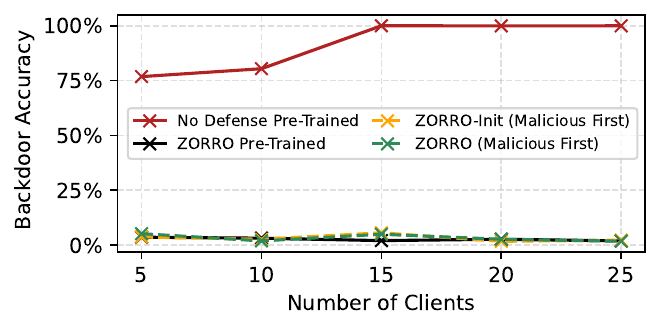}
	}
	
    \caption{\changedRevision{2}{BA for different client counts and initializations, including scenarios where malicious clients train first.}}
	\label{fig:eval-init}
\end{figure}

\subsection{Data Settings}
\label{sec:eval-setting}

\changedRevision{2}{Fig.~\ref{fig:eval-clients} visualizes the effectiveness of \ourname across different numbers of clients. As the figure shows, while BA under no defense remains consistently high, \ourname maintains a low BA across all settings, even in cases with a high client count, such as 1000 clients. \ourname remains effective, maintaining the MA stable, except for a small drop in the case of 1000 clients in both scenarios.}

\changedRevision{2}{In Fig.~\ref{fig:eval-init} we compare \ourname in scenarios where training starts with malicious clients, both with and without benign pre-training. To address the more challenging case without pre-training and adversarial client order, we propose a novel initialization method (see \sect~\ref{sec:sys-init}). As Fig.~\ref{fig:eval-init} shows, while \ourname alone can effectively remove malicious clients and maintain a low BA, the novel initialization is critical for handling edge cases as it is independent of client order and resilient to early-stage poisoning. In all experiments, it consistently selected a benign client as the entry point.}

Tab.~\ref{tab:eval-results:iid} shows the performance of \ourname under varying degrees of IID among clients. Across all settings, \ourname significantly reduces BA compared to No Defense and closely tracks the performance of the \goldenDefense. While MA increases with higher IID degrees for all defense schemes, most likely due to the training data becoming more representative, \ourname maintains consistently low BA even in highly non-IID scenarios, demonstrating its robustness in settings with heterogeneous data.

Fig.~\ref{fig:eval-pmr} shows the performance of \ourname for different PMRs, including PMR$=0\%$. As the figure shows, \ourname is also capable of achieving a decent accuracy in the absence of an attack, as it does not enforce discarding models if no attack is detected. The effectiveness for varying PMRs will be further discussed in \sect\ref{sec:eval-attacks}.

\begin{table}[b]
    \caption{\changedEditorial{Effectiveness of \ourname for different IID degrees in terms of Backdoor Accuracy (BA) and Main Task Accuracy (MA).}}
    \label{tab:eval-results:iid}
    \centering
    \scaleTable{\begin{tabular}{l|cc|cc|cc}
	IID & \multicolumn{2}{c|}{No Defense} & \multicolumn{2}{c|}{\goldenDefense} & \multicolumn{2}{c}{\ourname} \\
	& MA & BA & MA & BA & MA & BA \\
	\hline
	0.4 & 17.87\% & 58.43\% & 36.24\% & 1.01\% & 36.21\% & 2.78\% \\
	0.6 & 55.70\% & 66.22\% & 65.52\% & 5.81\% & 65.21\% & 4.30\% \\
	0.8 & 70.95\% & 80.41\% & 73.75\% & 2.82\% & 73.51\% & 3.16\% \\
	1.0 & 73.24\% & 88.80\% & 74.57\% & 5.33\% & 74.43\% & 4.61\% \\
\end{tabular}}
\end{table}

\subsection{Attack Settings}
\label{sec:eval-attacks}

\begin{figure}[t]
	\centering{	
			\includegraphics[width=0.775\columnwidth]{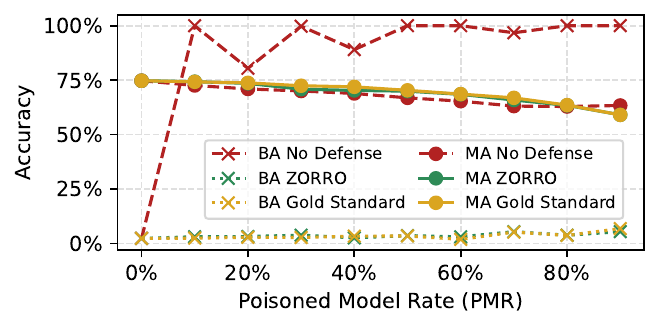}
	}
	\caption{\changedRevision{6}{Effectiveness of \ourname for different ratios of poisoned models (PMR).}}
	\label{fig:eval-pmr}
\end{figure}

Fig.~\ref{fig:eval-pmr} shows the effectiveness of \ourname under varying poisoning model ratios (PMR). As expected, BA under no defense remains consistently high across all PMR levels, while \ourname maintains a low BA close to the \goldenDefense baseline even as the number of malicious clients increases. Since with increasing PMR less benign clients are involved and, as such, also less benign data contribute, the MA decreases for all settings.

\begin{table}[t]

    \caption{Effectiveness of \ourname against various adaptive attack strategies, including Differential Privacy (DP)-based disguise, loss-constrained frequency optimization, varying poisoned data rates (PDRs)~\cite{baruch}, learning rate manipulations and Out Of Distribution samples (OOD)~\cite{wang2020attack}.}

	\label{tab:eval-results:attacks}
	\centering{	
            \scaleTable{
			\begin{tabular}{l|l|cccc}
	Defense & Attack & MA & BA & PRR & BBR \\
	\hline
	\goldenDefense & Default Attack & 73.75\% & 2.82\% & 100.00\% & 0.00\% \\
	\hline
	\multirow{8}{*}{\ourname}
	& DP Attack ($\epsilon{=}3.0$, $\delta{=}1.4$) & 73.02\% & 6.20\% & 100.00\% & 7.29\% \\
	& Frequency Attack & 73.51\% & 3.16\% & 94.66\% & 3.01\% \\
	& \changedRevision{3}{Low PDR (25\%)~\cite{baruch}} & 73.51\% & 3.16\% & 88.17\% & 3.01\% \\
	& \changedRevision{3}{Low PDR (50\%)~\cite{baruch}} & 73.51\% & 3.16\% & 94.17\% & 3.01\% \\
	& Default Attack & 73.51\% & 3.16\% & 94.66\% & 3.01\% \\
	& High PDR (100\%) & 73.51\% & 3.16\% & 98.87\% & 3.01\% \\
	& lr=0.0001 & 73.50\% & 3.35\% & 84.96\% & 3.79\% \\
	& lr=0.01 & 73.51\% & 3.16\% & 99.71\% & 3.01\% \\
    & \changedRevision{3}{OOD~\cite{wang2020attack}} & 73.29\% & 1.27\% & 90.55\% & 4.80\% \\
\end{tabular}
	}}
\end{table}

Tab.~\ref{tab:eval-results:attacks} evaluates the robustness of \ourname against various adaptive attack strategies. These include attempts to disguise backdoors using Differential Privacy, targeted manipulation of the frequency spectrum, and learning rate variations. As the results show, \ourname maintains strong defense performance across all scenarios, consistently keeping BA below $4\%$ and MA around $73.5\%$. \changedEditorial{Even under aggressive conditions such as full poisoning (PDR 100\%) or with learning rate manipulations, \ourname achieves high PRR and low BBR}. \changedRevision{3}{Lastly, we adapted existing Federated Learning attacks into the Split Learning paradigm. Specifically, we implemented the A-Little-Is-Enough attack~\cite{baruch} using reduced PDRs, and enhanced the Out-of-Distribution (OOD) attack with Projected Gradient Descent~\cite{wang2020attack}. More details are available in \app~\ref{app:eval-adaptive-attacks}.}

\subsection{Comparison with other Defenses}

\begin{table}[b]
    \caption{Comparison of \ourname with different baselines in terms of Backdoor Accuracy (BA), Main Task Accuracy (MA), Poisoned Removal Rate (PRR), and Better Benign Rate (BBR).}
    \label{tab:eval-results:defenses}
    \centering
    \scaleTable{\begin{tabular}{l|cccc}
	Defense & MA & BA & PRR & BBR\\\hline
	No Defense & 70.95\% & 80.41\% & \multicolumn{1}{c}{-} & \multicolumn{1}{c}{-} \\
	Differential Privacy & 65.70\% & 50.38\% & \multicolumn{1}{c}{-} & \multicolumn{1}{c}{-} \\
	Frequency Analysis & 61.51\% & 12.22\% & 54.64\% & 86.37\% \\
	\goldenDefense & 73.75\% & 2.82\% & 100.00\% & 0.00\% \\
	k-means & 52.06\% & 4.66\% & 100.00\% & 87.93\% \\\hline
	\ourname & 73.51\% & 3.16\% & 94.66\% & 3.01\% \\
\end{tabular}}
\end{table}

Tab.~\ref{tab:eval-results:defenses} compares \ourname against various baseline defenses. Particularly, we use Differential Privacy that clips the models at 1.4 and adds Gaussian noise with a standard deviation of 0.01. Further, we adapted SafeSplit~\cite{rieger25safesplit} and calculated the pairwise distances between the frequencies of the models in the current queue. In addition, we used k-means to spot anomalous model updates. However, as Tab.~\ref{tab:eval-results:defenses} shows, except for the \goldenDefense, only \ourname achieves a strong balance between security and utility, with a low BA of $3.16\%$, a high MA of $73.51\%$, and a PRR of $94.66\%$. In contrast, Differential Privacy and Frequency Analysis show either weak backdoor mitigation or significant drops in MA. While k-means achieves high PRR and low BA, its MA is considerably reduced to $52.06\%$, and it suffers from a high BBR of $87.93\%$, indicating that it frequently disregards newer benign models. Overall, \ourname outperforms prior defenses by combining strong robustness with practical training performance.

\subsection{Runtime Evaluation}
\label{sec:eval-zkpruntime}

\begin{figure}[tb]
	\centering{	
			\includegraphics[width=0.95\columnwidth]{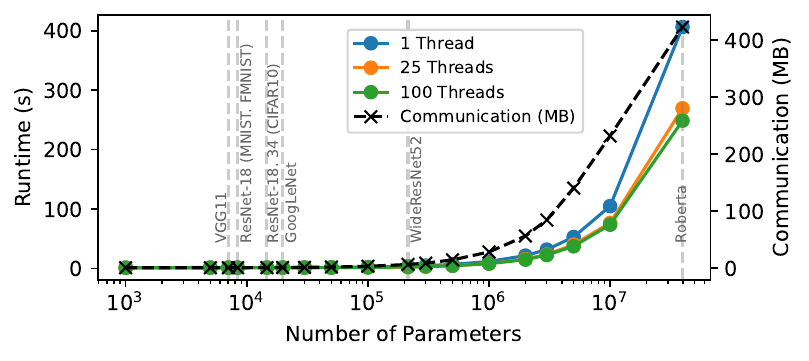}
	}
	\caption{\changed{Scalability of runtime and communication overhead of \ourname}}
	\label{fig:eval-zkpruntime}
\end{figure}

We evaluated the scalability of \ourname in terms of runtime and communication overhead. Fig.~\ref{fig:eval-zkpruntime} shows the prover's runtime for different numbers of threads (1, 25, and 100), alongside the communication cost in megabytes. As shown, the runtime scales sublinearly with the number of parameters for all thread configurations, demonstrating the benefits of parallelization. Communication overhead increases linearly but remains below 200\,MB even for models with up to $10^7$ parameters. These results demonstrate the scalability and efficiency of \ourname, even for large client-side model partitions.

\subsection{Memory Overhead of \ourname}
\label{sec:eval-edge}
\begin{table}[b]
    \caption{RAM Measurements for \ournameGen execution}
    \centering
    \scaleTable{
     \begin{tabular}{cc|rrr}
    \hline
    \textbf{Model} & \textbf{Dataset} & \multicolumn{1}{c}{\cellbreaks{\textbf{Client}\\\textbf{Parameters}}} & \multicolumn{1}{c}{\cellbreaks{\textbf{Max.}\\\textbf{RAM (MB)}}} & \multicolumn{1}{c}{\cellbreaks{\textbf{AVG.}\\\textbf{RAM (MB)}}} \\ \hline
    VGG11 & CIFAR-10 & 7050 & 636.91 & 449.97\\ \hline
    WideResNet52 & CIFAR-10 & 214436 & 823.45 & 553.10 \\ \hline
    WideResNet52 & Tiny ImageNet & 419336 & 942.77 & 666.63
    \\ \hline
    \end{tabular}
    }
    \label{tab:ram_analysis}
\end{table}

\changedRevision{4}{We show the RAM usage for the models with the smallest (VGG11) and largest (WideResNet52) amount of head and tail parameters in Tab.~\ref{tab:ram_analysis}. As can be seen, for the largest model, WideResNet52 trained on Tiny ImageNet, \ourname only requires at most 942.77 MB of RAM to ensure robustness during SL. Although ZKPs are generally memory-intensive, our meticulous design, paired with the fact that SL reduces the amount of parameters a client processes, ensures that any client with an edge device capable of vanilla SL can ensure security and robustness with \ourname with very little overhead. For instance, executing SL with WideResNet52 on a Raspberry Pi requires only 859.5~MB of RAM (median over 10 runs). Since the training process completes before ZKPs are executed, memory used for training becomes available again, rendering the additional memory overhead effectively negligible.}

\section{Security Analysis}
\subsection{Security of the ZKP Protocol}
\label{sec:security-zkp}
The security of \ourname against malicious clients attempting to subvert the backdoor defense critically relies on the cryptographic guarantees of the interactive zero-knowledge proof (ZKP) protocol, as described in \sect\ref{sec:sys-zkp}. We focus our analysis on the two fundamental properties of the ZKP: (1) \textit{computational soundness}, which ensures that no adversary can produce a valid proof for a false claim, and (2) \textit{zero-knowledge}, which guarantees that private inputs remain confidential. Together, these properties ensure the fulfillment of challenge C3 outlined in \sect\ref{sec:problem-requirements}.

Let \Prv denote the prover (client $C_i$), and \Vrf the verifiers (client $C_{i+1}$ and the server). Let $\Phi$ denote the statement being proven, namely: \Prv has correctly executed the defense protocol (Alg.~\ref{alg:zkp_resized}) using private model inputs $[M_{i-k}, \ldots, M_i]$ consistent with their public commitments and discrete cosine transforms, resulting in an updated top-$k$ model list forwarded to $C_{i+1}$. Let $pub$ and $priv$ denote the public and private inputs to the ZKP protocol, respectively, and let $\pi_i$ be the generated proof.

\subsubsection{Preventing Proof Forgery}
The central security guarantee is \textit{computational soundness}. The Wolverine ZKP protocol~\cite{weng2021wolverine} ensures that a polynomial-time bounded malicious prover $\mathcal{P}^*$ cannot convince an honest verifier \Vrf to accept a proof $\pi^*$ for a false statement $\Phi$, except with negligible probability $\varepsilon$, inherent to the VOLE-based construction. Formally, let $\mathcal{L}$ be the language of true statements such that $(pub, priv) \in \mathcal{L}$ iff $\Phi(pub, priv)$ is true. Then:

\begin{equation}
    \forall \text{ PPT } P^*, Pr[\mathcal{V} \text{ accepts } \pi^* \ \text{from} \ P^*(pub):(pub,\cdot)\notin\mathcal{L}] \leq \varepsilon
\end{equation}

\ourname's use of interactive ZKPs guarantees that a malicious client $C_i$ cannot generate a valid proof if it deviates from the defense protocol. Examples of invalid behaviors include:
\begin{itemize}[leftmargin=*]
    \item \textit{Manipulating poison risk scores}: Falsifying $S_t^*$ to misclassify a poisoned model as benign.
    \item \textit{Providing inconsistent inputs}: Using model parameters not matching their public commitments.
    \item \textit{Tampering with pruning logic}: Incorrectly updating the top-$k$ model list by retaining a high-risk model.
\end{itemize}

Any such deviation leads to a failed verification of $\pi^*$, resulting in rejection by honest verifiers with probability at least $1-\varepsilon$. In scenarios where a malicious verifier colludes with the prover, \ourname mitigates this risk by requiring all proofs to be sent to the semi-honest server, which can terminate the protocol if a violation is detected. Thus, clients cannot successfully forge proofs without detection, ensuring the integrity of the defense.

\subsubsection{Preserving Model Privacy}
The \textit{zero-knowledge} property ensures that the ZKP reveals no information about the prover’s private inputs beyond the validity of the claim. Formally, for any probabilistic polynomial-time verifier $\mathcal{V}^*$, there exists a simulator $Sim$ such that the verifier's view in the real protocol is computationally indistinguishable from the simulator's output using only the public inputs $pub$ and the truth of $\Phi$:

\begin{equation}
\begin{aligned}
    & {View_{V*}(P(priv, pub) \leftrightarrow V^*(pub))}_{(pub, priv)} \simeq  \\ &{Sim(pub, IsTrue(\Phi(pub, \cdot)))}_{(pub, priv)}
\end{aligned}
\end{equation}

Here, $\text{View}_{\mathcal{V}^*}$ comprises all messages exchanged during proof generation and verification. This ensures that neither the server nor $C_{i+1}$, acting as verifiers, learn any sensitive information such as the model parameters of $C_i$. Furthermore, since only clients exchange top-$k$ model partitions, the server never receives raw model parameters, satisfying Requirement R3 in \sect\ref{sec:problem-requirements}.
\subsubsection{Completeness}
The \textit{completeness} property guarantees that any honest prover following the protocol with valid private inputs will have its proof $\pi_i$ accepted by the verifiers with high probability ($1 - \varepsilon$). This ensures that correct behavior is not penalized and that the system remains functional even under strict verification.\\

The VOLE-based ZKP construction used in \ourname provides rigorous cryptographic guarantees of soundness, completeness, and zero-knowledge. These properties ensure that malicious clients cannot forge their way past the defense protocol, while honest clients can prove compliance without compromising model privacy. Moreover, by incorporating server verification and leveraging the security parameter $k$, \ourname remains resilient against collusion among multiple malicious clients, preserving both the integrity and privacy of the collaborative training process.

\subsection{Security of Poisoning Detection}
\label{sec:security-poisoning}
The security of \ournameGen poisoning detection relies on the integrity of the client-side defense scheme, which is cryptographically enforced through interactive zero-knowledge proofs (ZKPs). As analyzed in detail in \sect\ref{sec:security-zkp}, the ZKPs ensure that clients correctly execute the defense protocol, including poison scoring, model pruning, and pointer updates, without revealing any private model information. This cryptographic attestation guarantees that even adversarial clients cannot deviate from the defense procedure, thereby ensuring protocol compliance and robustness against manipulation. The security of the ZKP protocol was proven in \sect\ref{sec:security-zkp}

\changed{An important invariant for \ournameGen security is that the model queue always contains at least one benign model, forming the inductive base case. In each round, exactly one new model is added and one is removed, preserving a constant queue size. Assuming the invariant holds at round~$i$, then at most one poisoned model may be introduced in round~$i+1$ by the current client, while one model, potentially poisoned, is discarded. If the added model shows detectable poisoning artifacts, the scoring function will prioritize its removal. Thus, if there was one benign model in the queue before the current round (as given by the security invariant) and the new model, which, if poisoned, is removed, it follows that after the defense round, again one benign model is in the queue, thus fulfilling the security invariant also for the following round. The analysis technique of \ourname allows choosing this benign model when it shows the lowest poisoned score, independent of the number of further poisoned models in the queue. The effectiveness of \ournameGen defense is extensively demonstrated in \sect\ref{sec:eval}, showing high poisoned removal rates, {low BA}, and strong resilience across varying datasets, model architectures, and attacker strategies.}
\section{Related Works}
Various schemes have been proposed in the past to mitigate backdoor attacks in distributed learning paradigms. In the following, we discuss existing defense mechanisms, schemes specific to SL (\sect\ref{sec:sota-sl}). \changedEditorial{We discuss recent approaches leveraging ZKPs to enhance the security of deep learning (\sect\ref{sec:sota-zkp}). Further, we analyze defenses for other distributed learning paradigms in App.~\ref{app:sota-distributed}).}

\subsection{Backdoor Defenses in Split Learning}
\label{sec:sota-sl}

Pu \etal~\cite{pu2024dullahan} investigate attacks initiated by malicious servers within SL. Although their primary contribution is an attack strategy, they also explore potential defense mechanisms. Specifically, they propose adding noise to the client-side training data to hinder the server’s ability to construct an accurate surrogate model. However, this defense is ineffective against client-side attacks, where adversarial clients can arbitrarily alter model components and inject manipulated parameters. 
Similarly, Bai \etal introduce an attack method but incorporate defense considerations in their robustness evaluation. They describe two mitigation strategies for the server, one involves analyzing embeddings to detect duplicates, and the other aims to smooth embeddings to reduce the impact of adversarial manipulations. In contrast, Rieger \etal~\cite{rieger25safesplit} assume a majority of benign clients and employ pairwise distance calculations to identify outliers. However, their approach requires storing a list of the most recent update from each client and does not allow discarding all but a small subset of client models, as enabled by \ourname, thus raising concerns about scalability and privacy. Importantly, such server-side defenses are complementary to \ourname and can be easily integrated to enable joint inspection of both server-side and client-side model partitions.

\subsection{Applications of ZKP for Secure Deep Learning}
\label{sec:sota-zkp}
While ZKPs have not been used to secure split learning, they have been used for both verifiable machine learning and defending learning paradigms, such as federated learning. Verifiable machine learning has grown to prominence as a technique to allow cloud service providers to attest to the computational integrity of outsourced inference tasks to consumers without revealing any proprietary information about their models. While the seminal works in this field \cite{chen2024zkml, sun2024zkllm, liu2021zkcnn} primarily focus on verifiable inference, there has been an emerging thread of research that focuses on providing zero-knowledge proofs of training (zkPoTs) \cite{liu2021zkcnn, garg2023experimenting}.

In the context of FL, ZKPs have been utilized to design privacy-preserving defenses that ensure both the integrity and confidentiality of client contributions, even in the presence of malicious clients. Notable works~\cite{ghodsi2023zprobe, lycklama2023rofl, roy2022eiffel} use ZKPs to prevent malformed updates from corrupting the global model. These schemes typically rely on aggregation-based statistics, allowing clients to demonstrate the validity of their updates without revealing sensitive training data. Although effective in FL, these approaches are not directly applicable to SL due to its sequential training structure, where aggregation-based defenses lose efficacy.

\ourname introduces the first ZKP-based approach for securing Split Learning. By incorporating interactive ZKPs, \ourname achieves both robustness and privacy in the presence of malicious clients aiming to inject backdoors while maintaining minimal computational overhead. Furthermore, the modular implementation of \ourname lays a foundation for future client-side defenses in SL. Researchers and developers can easily extend \ourname by integrating new defense modules, such as clustering, filtering, or norm-based metrics, within the existing framework, facilitating the development of verifiable and privacy-preserving defense strategies for Split Learning.

\section{Conclusion}

The distributed nature of SL makes it vulnerable to backdoor attacks, particularly from malicious clients that can inject poisoned updates during collaborative training. We presented \ourname, a novel client-side defense mechanism that ensures robustness and privacy in SL through the use of interactive ZKP. \ourname outsources not only the training but also the defense itself, enabling a comprehensive analysis of the client-located model partitions. By leveraging our novel interactive ZKP protocol, we ensure the correct execution of the defense and compel malicious clients to expose themselves, allowing subsequent clients to safely bypass poisoned training contributions. Our poison risk scoring approach analyzes model updates in the frequency domain, enabling accurate backdoor detection without relying on assumptions about the attack strategy or the proportion of malicious clients.

\begin{acks}
This research received funding from the Horizon program of the European Union under grant agreements No. 101093126 (ACES) and No. 101070537 (CROSSCON), as well as the Federal Ministry of Education and Research of Germany (BMBF) within the IoTGuard project, and the Deutsche Forschungsgemeinschaft (DFG) SFB-1119 CROSSING/236615297. This work is also supported by the U.S. Army/Department of Defense award number W911NF2020267 and the Defense Advanced Research Projects Agency (DARPA) Proofs program under Grant No. HR0011-23-1-0006.
\end{acks}

\bibliographystyle{ACM-Reference-Format}
\balance
\bibliography{main}

\appendix
\section{Frequency Selection}
\label{app:low-frequencies}

\changedEditorial{For the frequency transformation, we employ the \textit{Discrete Cosine Transform} (DCT). The DCT has been proven to closely approximate the \textit{Karhunen-Lo\`eve Transform} (KLT)~\cite{li2019energy}, which is theoretically optimal for energy compaction~\cite{dur1998optimality}. In contrast to the Fast Fourier Transform (FFT), which yields complex-valued outputs, the DCT operates entirely in the real domain, simplifying both computation and interpretation. Moreover, due to its strong energy compaction properties, the DCT is particularly effective at emphasizing subtle, systematic deviations, such as those introduced by backdoor attacks, in a small set of low-frequency components. This makes it especially well-suited for detecting poisoned artifacts in model updates, while maintaining computational efficiency and robustness in security-sensitive environments.}

\ourname uses the low-frequency components for analyzing the model updates for poisoned artifacts.
Let \( X(u,v) \) be the 2D DCT coefficients of an \( N \times N \) block, where \( u \) and \( v \) are the vertical and horizontal frequency indices. The selected low-frequency coefficients form an upper-left triangular region:
\begin{equation}
\{ X(u,v) \mid 0 \leq v < N/2,\ 0 \leq u < N/2 - v \},
\end{equation}
corresponding to the case \( u + v < N/2 \).

\section{Datasets}
\label{app:datasets}
\changedEditorial{We split a distributed setting by partitioning the training dataset. To simulate different degrees of similarity of the data distributions (independent and identically distributions, IID), \changedEditorial{we} follow the main label strategy, aligned with existing work on distributed machine learning~\cite{cao2021fltrust,rieger25safesplit}. For each client a main label is randomly selected. A fraction of the dataset that is equivalent to the IID-degree is drawn randomly from all training samples while the remaining part is drawn only from samples of the main label.}\\

For our evaluation, we use three image datasets that are frequently used for deep learning and SL in particular~\cite{rieger25safesplit,tajalli2023feasibility}.\\
\textbf{\cifar}  is a widely used image classification dataset consisting of \numprint{60000} color images divided into 10 classes, such as airplanes, automobiles, birds, and ships. Each image has a resolution of $32 \times 32$ pixels and contains three color channels (RGB). The dataset is split into \numprint{50000} training samples and \numprint{10000} test samples~\cite{krizhevsky2009learning}. The backdoor trigger is a red rectangle in the upper left corner, making the sample classified as 'bird'.

\noindent\textbf{Modified National Institute of Standards and Technology (\mnist)} is a benchmark dataset of handwritten digit images, containing \numprint{70000} grayscale images categorized into 10 digit classes (0–9). Each image is $28 \times 28$ pixels in size and contains a single channel (grayscale). The dataset is split into \numprint{60000} training samples and \numprint{10000} test samples. MNIST is known for its simplicity~\cite{deng2012mnist}. The backdoor trigger is a white rectangle in the upper left corner, making the sample classified as a '3'.

\noindent\textbf{Fashion-MNIST (\fmnist)} is a drop-in replacement for MNIST that contains \numprint{70000} grayscale images of fashion items, such as shirts, shoes, and bags, across 10 classes. Like MNIST, each image is $28 \times 28$ pixels with a single grayscale channel. The dataset is similarly partitioned into \numprint{60000} training and \numprint{10000} test samples. Fashion-MNIST provides a slightly more complex and realistic alternative to MNIST while maintaining the same structure and size, making it suitable for benchmarking learning algorithms under consistent conditions~\cite{fmnist}. The backdoor trigger is a white rectangle in the upper left corner, making the sample classified as 'dress'.

\noindent\changedRevision{5}{\textbf{\cifarhundred} a dataset similar to \cifar, it consists of \numprint{60000} color images divided into 20 superclasses, each consisting of 5 classes. Each image has a resolution of $32 \times 32$ pixels and contains three color channels (RGB). The dataset is split into \numprint{50000} training samples and \numprint{10000} test samples~\cite{krizhevsky2009learning}. We trained \cifarhundred with 100 clients, and each with a unique main label for the non-IID experiments. The backdoor trigger is a red rectangle in the upper left corner, which makes the sample misclassified as 'chair'.}

\noindent\changedRevision{5}{\textbf{\tinyimagenet} is a subset of the ImageNet~\cite{imagenet-object-localization-challenge} dataset consisting of 200 classes~\cite{tiny-imagenet}. It contains \numprint{100000} training samples and \numprint{10000} test samples, all with a resolution of $64\times 64$ pixels containing three color channels (RGB). The dataset is downloaded from \url{https://huggingface.co/datasets/zh-plus/tiny-imagenet}. Similar to \cifarhundred, we trained our models with 100 clients, but 2 main labels per client. The backdoor trigger is a red rectangle in the upper right corner, making the sample misclassify as 'bullfrog'.}

\noindent\changedRevision{5}{\textbf{\forest} is a tabular dataset that contains tree observations from areas of the Roosevelt National Forest in Colorado~\cite{covertype_31}. It consists of 54 features and 7 classes. We dropped classes 3 and 4 due to a low number of samples compared to the other classes. Therefore, our dataset consists of \numprint{398140} training and \numprint{170632} test samples. For training this tabular dataset, we created a simple feed-forward neural network with \numprint{589541} parameters and 10 linear layers stacked with batch normalization, ReLU activation, and dropout. To train the model, we used in all cases a PDR of 50\%. The backdoor shall be activated if the feature elevation is bigger than \numprint{3500}, making the sample classified as 'aspen'.}

\FloatBarrier
\section{Variance of Experiment Results}
\label{app:scattering}
\begin{figure}[t]
	\centering{	
			\includegraphics[width=0.65\columnwidth, trim=0cm 0.25cm 0cm 0cm, clip]{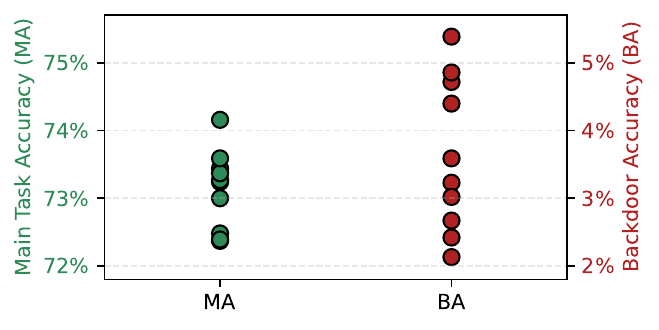}
	}
	\caption{Variance in the BA and MA for \ourname.}
	\label{fig:eval-scattering}
\end{figure}
To assess the statistical robustness of the results, we conducted an experiment on each of the used ML servers 5 times with different seeds. As Fig.~\ref{fig:eval-scattering} shows, the values show only a small variance with 0.55\% (MA) and 1\% (BA).

\section{Ablation Study}
\label{app:ablation}

\begin{table}[H]
    \caption{Effectiveness of different \queueLength values in terms of Backdoor Accuracy (BA), Main Task Accuracy (MA), Poisoned Removal Rate (PRR), and Better Benign Rate (BBR).}
    \label{tab:eval-results:ql}
    \centering
    \scaleTable{\begin{tabular}{lr|cccc}
	Defense & \queueLength & BA & MA & PRR & BBR\\\hline
	No Defense &  & 80.41\% & 70.95\% & \multicolumn{1}{c}{-} & \multicolumn{1}{c}{-} \\
	\goldenDefense &  & 2.82\% & 73.75\% & 100.00\% & 0.00\% \\\hline
	\multirow{16}{*}{\ourname} & 1 & 35.43\% & 73.10\% & 87.55\% & 0.00\% \\
	& 2 & 2.82\% & 73.75\% & 91.91\% & 0.00\% \\
	& 3 & 3.16\% & 73.51\% & 94.66\% & 3.01\% \\
	& 4 & 3.46\% & 72.45\% & 91.26\% & 9.18\% \\
	& 5 & 3.24\% & 72.58\% & 91.71\% & 9.52\% \\
	& 6 & 3.55\% & 71.41\% & 92.30\% & 11.85\% \\
	& 7 & 3.33\% & 71.40\% & 89.36\% & 11.58\% \\
	& 8 & 4.72\% & 71.08\% & 79.62\% & 12.27\% \\
	& 9 & 4.62\% & 69.98\% & 88.88\% & 11.86\% \\
	& 10 & 4.99\% & 69.91\% & 80.98\% & 13.19\% \\
	& 15 & 10.31\% & 68.31\% & 76.78\% & 15.53\% \\
	& 20 & 5.61\% & 68.54\% & 89.99\% & 17.28\% \\
	& 25 & 8.56\% & 66.08\% & 85.68\% & 16.78\% \\
	& 30 & 5.21\% & 64.97\% & 87.44\% & 17.03\% \\
	& 35 & 6.07\% & 60.36\% & 90.95\% & 18.20\% \\
	& 40 & 4.84\% & 58.99\% & 91.68\% & 17.61\% \\
\end{tabular}}
\end{table}

We evaluated the impact of different choices for the security parameter \betaParam on the performance of \ourname. As Tab.~\ref{tab:eval-results:beta} shows, a smaller value slightly reduces the BBR, making it more likely to select the most recent model. However, this also increases the risk of selecting a poisoned model. For $\betaParam = 0.5$, the PRR drops significantly and the BA rises to $34.94\%$. Conversely, setting $\betaParam = 1.0$ ensures that poisoned models are never selected, as their risk scores are consistently higher than those of benign models. Yet, this strictness causes the most recent benign model to be ignored in $89.73\%$ of cases, which significantly impairs training and leads to a substantial drop in MA. In contrast, $\betaParam = 0.7$ strikes a practical balance between utility and robustness, reducing BA to the level of the \goldenDefense while preserving high MA.

Tab.~\ref{tab:eval-results:distances} shows the results of our ablation study for different choices for distance metrics for poison scoring. As the results show, using the cosine distance fails to detect poisoned models effectively, yielding BA and MA identical to the no-defense baseline. Switching to $L_2$-norm significantly improves performance, reducing BA to $2.65\%$ and achieving a high PRR of $94.95\%$. However, this comes at a slight cost to BBR. The default choice in \ourname, the Taxicab norm, offers a balanced trade-off, maintaining low BA ($3.16\%$), high MA ($73.51\%$), and strong PRR ($94.66\%$), while also keeping BBR lower than $L_2$-norm. This confirms the effectiveness of the Taxicab-based scoring in balancing robustness and training utility.

Tab.~\ref{tab:eval-results:ql} evaluates the impact of the queue length \queueLength on the performance of \ourname. From a privacy and computational efficiency perspective, it is desirable to keep \queueLength as small as possible, since fewer models need to be exchanged, stored, and evaluated at each client and also less other clients gain access to the clients' models. However, a too small \queueLength (e.g., \queueLength$=1$) poses a security risk, as it reduces the context for identifying poisoned models and increases the chance that a malicious client can evade detection. On the other hand, a large \queueLength (e.g., \queueLength$\geq 20$) may result in the selection of outdated models, which hinders training progress and leads to reduced MA and higher BBR. As the table shows, a moderate value of \queueLength$=3$ achieves the best balance, minimizing BA ($3.16\%$), maintaining high MA ($73.51\%$), and ensuring robust PRR and low BBR.

In App.~\ref{app:distances}, we exemplary plot the poison-risk scores for one repetition of \cifar.

\begin{table}[b]
    \caption{Effectiveness of different \betaParam values in terms of Backdoor Accuracy (BA), Main Task Accuracy (MA), Poisoned Removal Rate (PRR), and Better Benign Rate (BBR).}
    \label{tab:eval-results:beta}
    \centering
    \scaleTable{\begin{tabular}{ll|cccc}
	Defense & \betaParam & BA & MA & PRR & BBR\\\hline
	No Defense &  & 80.41\% & 70.95\% & \multicolumn{1}{c}{-} & \multicolumn{1}{c}{-} \\
	\goldenDefense &  & 2.82\% & 73.75\% & 100.00\% & 0.00\% \\\hline
	\multirow{3}{*}{\ourname} & 0.5 & 34.94\% & 73.48\% & 54.33\% & 3.12\% \\
	& 0.7 & 3.16\% & 73.51\% & 94.66\% & 3.01\% \\
	& 1.0 & 0.00\% & 23.58\% & 100.00\% & 89.73\% \\
\end{tabular}}
\end{table}
\begin{table}[bt]
    \caption{Ablation study of \ourname using different distance metrics in terms of Backdoor Accuracy (BA), Main Task Accuracy (MA), Poisoned Removal Rate (PRR), and Better Benign Rate (BBR).}
    \label{tab:eval-results:distances}
    \centering
    \scaleTable{\begin{tabular}{ll|cccc}
	Defense & Distance-Metric & BA & MA & PRR & BBR\\\hline
	No Defense &  & 80.41\% & 70.95\% & \multicolumn{1}{c}{-} & \multicolumn{1}{c}{-} \\
	\goldenDefense &  & 2.82\% & 73.75\% & 100.00\% & 0.00\% \\\hline
	\multirow{3}{*}{\ourname} & Cosine & 80.41\% & 70.95\% & 31.06\% & 0.00\% \\
	& $L_2$-Norm & 2.65\% & 73.39\% & 94.95\% & 4.17\% \\
	& Taxicab (\ourname) & 3.16\% & 73.51\% & 94.66\% & 3.01\% \\
\end{tabular}}
\end{table}

\section{Distance Values}
\label{app:distances}

\begin{figure}[t]
	\centering{	
			\includegraphics[width=0.775\columnwidth]{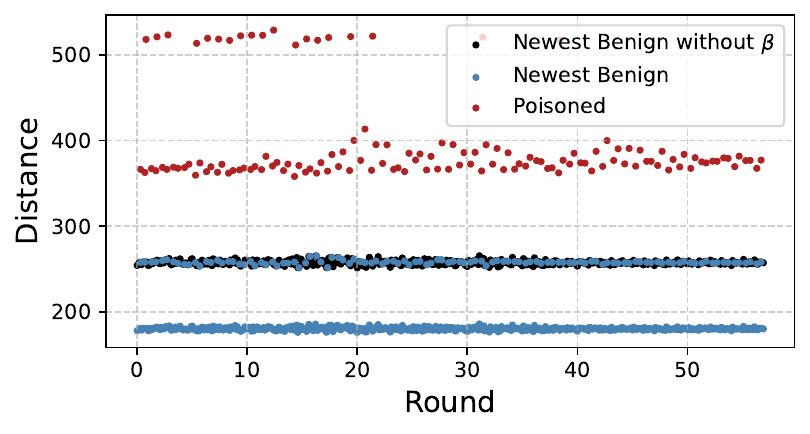}
	}
    	\caption{Poison risk-scores for benign and poisoned model updates.}
	\label{fig:eval-distances}
\end{figure}

Figure~\ref{fig:eval-distances} exemplarily visualizes the poison risk scores assigned to poisoned models, the most recent benign model, and the most recent benign model without applying the security parameter $\betaParam$. For illustration purposes, the figure presents results from a single random seed. The visualization shows the effectiveness of the poison scoring function in distinguishing between benign and poisoned models. As shown, poisoned models consistently receive significantly higher risk scores than benign models, even when $\betaParam$ is not applied. 

Notably, some benign model scores (in blue) are identical to the scores obtained without applying $\betaParam$. These instances correspond to rounds in which a poisoned model was most recently added. Consequently, the $\betaParam$ adjustment was applied to the poisoned model rather than the benign one, resulting in unscaled benign scores. This behavior further emphasizes the discriminative capability of the scoring function in practical settings.

\section{Distraction of Benign Model for \mnist}
\label{app:mnistAnomaly}

In Table~\ref{tab:eval-results:datasets}, we observed that the BA reached 100\% in one experiment repetition on the \mnist dataset, although all poisoned models were immediately excluded (PRR = \numprint{100}\%). We hypothesized that the presence of the trigger might unintentionally distract the benign model, causing it to consistently predict a specific label when facing to triggered inputs. To test this hypothesis, we conducted an additional experiment in which we trained three benign models (PMR = \numprint{0}\%) with different random seeds. For each model, we created ten separate test sets, each containing only one label, while applying the same trigger (a white rectangle in the upper-left corner) to every input.

Interestingly, although none of these models were ever trained on poisoned data, each benign model consistently predicted a single label for all triggered samples, regardless of the true class. Consequently, if this randomly predicted label happened to match the backdoor target label, the resulting BA would be 100\% despite the absence of poisoning. Notably, the consistently predicted label varied across models trained with different seeds.

This behavior is most likely caused by overfitting to the structural biases of the \mnist dataset. Since the dataset is highly homogeneous, with digits always centered and the surrounding regions consistently black, the models rarely, if ever, learn to process input variations in the outer image regions. As a result, the sudden appearance of a trigger in those regions produces undefined behavior, which manifests as a consistent, yet arbitrary prediction. However, in contrast to actual backdoor attacks, where the attacker can choose a specific target label, this effect is random and non-controllable by the adversary. Notably, we repeated this experiment for \cifar but this phenomenon was not observed here, as the samples in \cifar are more heterogeneous and vary in size and position.

\section{Backdoor Defenses in Other Distributed Learning Paradigms}
\label{app:sota-distributed}
Federated learning (FL)~\cite{mcmahan2017} represents another distributed learning paradigm that enables clients to collaboratively train a deep neural network without sharing raw data. FL has been extensively studied from a security perspective. In this setting, clients train locally using a shared model checkpoint and subsequently send their updates to a central server, which aggregates them and redistributes the updated model for further training iterations.

Defense mechanisms in FL can be broadly categorized into active approaches, which aim to identify malicious clients~\cite{shen16Auror,fung2020FoolsGold,krauss2023mesas}, and passive approaches, which modify the aggregation rule~\cite{blanchard17Krum,yin2018Median} or apply smoothing techniques to mitigate potential backdoors~\cite{bagdasaryan, naseri2022local}. However, active methods typically rely on comparing model updates across clients to detect anomalies~\cite{shen16Auror,fung2020FoolsGold,krauss2023mesas}, which is infeasible in SL due to its sequential training structure (see C2 in \sect\ref{sec:problem-requirements}). Also, aggregation-based defenses~\cite{blanchard17Krum,yin2018Median} are not applicable, as SL does not involve aggregating model updates. On the other side, smoothing-based defenses~\cite{naseri2022local,bagdasaryan} may degrade model performance by damaging learned representations.

In contrast, \ourname leverages the frequency-domain analysis of model representations to effectively detect poisoned models, even when they originate from different checkpoints. This enables robust detection in scenarios unique to SL, where traditional FL defenses fall short.

\section{\ournameGen Effectiveness on adapted Attack from Other Distributed Learning Paradigms}
\label{app:eval-adaptive-attacks}

\changedRevision{3}{We adapted the A-Little-Is-Enough attack by Baruch et al.~\cite{baruch} to the sequential setting of Split Learning. By injecting small poisoned data rates (PDRs), we achieved a backdoor accuracy of 78\% and 81\% for 25\% and 50\% PDR, respectively, in the No Defense case. However, when applying \ourname, the attack is effectively mitigated: the malicious client is removed in over 85\% of the cases, preventing it from embedding a meaningful backdoor into the global model.

In addition, we implemented the Attack on the Tail backdoor from Wang et al.~\cite{wang2020attack}, which does not rely on explicit triggers or artifacts. Instead, it targets out-of-distribution (OOD) samples, for instance, misclassifying images of Southwest airplanes as trucks. The attack exploits the assumption that benign clients lack samples from the target distribution, a condition we replicate using the authors’ original code. The attack is further enhanced with the use of Projected Gradient Descent optimization and achieves a backdoor accuracy of 95\% on the \cifar dataset.

When the system employs \ournameGen update scoring and immediate pruning, we observed that poisoned models are effectively filtered out, maintaining robustness against adaptive attacks. The nature of Split Learning, where updates are applied sequentially, further supports this defense by allowing benign client contributions to overwrite minor manipulations, while \ourname detects and removes larger ones. Tab.~\ref{tab:eval-results:attacks} confirms that backdoor accuracy remained low (3.16\%, 1.27\%) even when detection scores slightly declined, highlighting \ournameGen robustness against adaptive adversarial behavior.}

\end{document}